\documentclass{aa}

\usepackage{amssymb}
\usepackage{amsmath}
\usepackage{graphicx}
\usepackage{natbib}
\usepackage[varg]{txfonts}
\usepackage{caption}
\usepackage{hyperref}
\usepackage{graphicx}
\usepackage{xcolor}
\usepackage[normalem]{ulem}

\hypersetup{
    colorlinks=true,
    linkcolor=blue,
    filecolor=magenta,
    urlcolor=blue,
    citecolor=blue,
}

\bibpunct{(}{)}{;}{a}{}{,}

\newcommand{\srcname}{J085039\xspace}
\newcommand{\fullname}{1eRASS J085039.9-421151\xspace}

\definecolor{myBlue}{rgb}{0,0.5,1}

\def\xsh{\textit{X-shooter}\xspace}

\defcitealias{2024MNRAS.528L..38D}{De+24}

\begin{document}

\title{Multiwavelength study of \fullname with eROSITA, \textit{NuSTAR} and \textit{X-shooter}}

\author{A.~Zainab\inst{\ref{inst:remeis}}\thanks{\email{aafiazainab.ansar@fau.de}}, A.~Avakyan\inst{\ref{inst:tue}}, V.~Doroshenko\inst{\ref{inst:tue}}, P.~Thalhammer\inst{\ref{inst:remeis}}, E.~Sokolova-Lapa\inst{\ref{inst:remeis}}, R.~Ballhausen\inst{\ref{inst:umcp},\ref{inst:godd}}, N.~Zalot\inst{\ref{inst:remeis}}, J.~Stierhof\inst{\ref{inst:remeis}}, S.~H\"ammerich\inst{\ref{inst:remeis}}, C.~M.~Diez\inst{\ref{inst:esac}}, P.~Weber\inst{\ref{inst:remeis}}, T.~Dauser\inst{\ref{inst:remeis}}, K.~Berger\inst{\ref{inst:remeis}}, P.~Kretschmar\inst{\ref{inst:esac}}, K.~Pottschmidt\inst{\ref{inst:csst},\ref{inst:godd}}, P.~Pradhan\inst{\ref{inst:erau}}, N.~Islam\inst{\ref{inst:csst},\ref{inst:godd}}, C.~Maitra\inst{\ref{inst:mpe}}, J.~B.~Coley\inst{\ref{inst:how},\ref{inst:godd}}, P.~Blay\inst{\ref{inst:esict},\ref{inst:tu-esac}}, R.~H.~D.~Corbet\inst{\ref{inst:csst},\ref{inst:godd}}, R.~E.~Rothschild\inst{\ref{inst:ucsd}}, K.~Wood\inst{\ref{inst:nrl}}, A.~Santangelo\inst{\ref{inst:tue}}, U.~Heber\inst{\ref{inst:remeis}}, J.~Wilms\inst{\ref{inst:remeis}}}

\authorrunning{A.~Zainab}

\institute{Dr.\ Karl-Remeis Sternwarte and Erlangen Centre for Astroparticle Physics, Friedrich-Alexander Universit\"at Erlangen-N\"urnberg, Sternwartstr.~7, 96049 Bamberg, Germany 
\label{inst:remeis}
\and
Universit\"at T\"ubingen, Institut f\"ur Astronomie und Astrophysik T\"ubingen, Sand 1, 72076 T\"ubingen, Germany
\label{inst:tue}
\and
University of Maryland College Park, Department of Astronomy, College Park, MD 20742, USA
\label{inst:umcp}\and
NASA Goddard Space Flight Center, Astrophysics Science Division, Greenbelt, MD 20771, USA
\label{inst:godd}
\and
European Space Agency (ESA), European Space Astronomy Centre (ESAC), Camino Bajo del Castillo s/n, 28692 Villanueva de la Cañada, Madrid, Spain
\label{inst:esac}
\and
Center for Space Science and Technology, University of Maryland Baltimore County, 1000 Hilltop Circle, Baltimore, MD 21250, USA
\label{inst:csst}
\and 
Embry Riddle Aeronautical University, Department of Physics, 3700 Willow Creek Road, Prescott, AZ, 86301, USA
\label{inst:erau}
\and
Max-Planck-Institut f\"ur extraterrestrische Physik, Giessenbachstraße 1, 85748 Garching, Germany
\label{inst:mpe}
\and
Department of Physics and Astronomy, Howard University, Washington, DC 20059, USA
\label{inst:how}
\and
Department of Astronomy and Astrophysics, University of California, San Diego, 9500 Gilman Dr., La Jolla, CA 92093-0424, USA
\label{inst:ucsd}
\and
Praxis Inc., Alexandria, VA 22303, resident at Naval Research Laboratory, Washington, DC 20375, USA
\label{inst:nrl}
\and
ESICT, Valencian International University, Pintor Sorolla 21, 46002 Valencia, Spain
\label{inst:esict}
\and
Telespazio UK for ESA, European Space Astronomy Center (ESAC), Camino bajo del castillo S/N, 28692 Villanueva de la Ca\~nada, Spain
\label{inst:tu-esac}
}

\abstract{The eROSITA instrument on board \textit{Spectrum-Roentgen-Gamma} has completed four scans of the
  X-ray sky, leading to the detection of almost one million X-ray
  sources in eRASS1 only, including multiple new X-ray binary candidates.
  We report on analysis of the X-ray binary \fullname, using a ${\sim}$55\,ks long
  \textit{NuSTAR} observation, following its detection in each eROSITA scan. Analysis of the eROSITA and
  \textit{NuSTAR} X-ray spectra in combination with \xsh data
  of the optical counterpart provide evidence of an X-ray binary with a red supergiant (RSG) companion, confirming previous results, although we determine a cooler spectral type of M2--3, owing to the presence of TiO bands in the optical and near infrared spectra.
  The X-ray spectrum is well-described by an absorbed power law with a
  high energy cutoff typically applied for accreting high mass X-ray binaries. In addition, we detect a strong fluorescent neutral iron line with an equivalent width of ${\sim}$700\,eV and an absorption edge, the latter indicating strong absorption by a partial covering component. It is unclear if the partial absorber is ionised. There is no significant evidence of 
  a cyclotron
  resonant scattering feature. We do not detect any pulsations in the
  \textit{NuSTAR} lightcurves, possibly on account of a large spin period
  that goes undetected due to insufficient statistics at low
  frequencies or potentially large absorption that causes pulsations to be smeared out. Even so, the low persistent luminosity, the spectral parameters observed (photon index, $\Gamma<1.0$), and the minuscule likelihood of detection of RSG-black hole systems, suggest that the compact object is a neutron
  star.}
    
\maketitle

\section{Introduction}

As part of its X-ray all-sky
survey, the extended ROentgen Survey Imaging Telescope Array \citep[eROSITA,][]{Merloni12, 2021A&A...647A...1P} on board
the Spectrum-Roentgen-Gamma (\textit{Spektr-RG}, \textit{SRG})
observatory has completed four surveys  (out of a planned eight), each with a duration
of six months. The eROSITA survey has already detected several million X-ray sources and has increased the number of known X-ray
sources by a significant factor compared to the 1990 sky survey
conducted by ROentgen SATellite \citep[\textit{ROSAT},][]{Voges99, 2016A&A...588A.103B}. 
Among other source types, eROSITA is expected to significantly increase the number of known X-ray binaries (XRBs), particularly those with lower luminosities \citep{2014A&A...567A...7D}. Indeed, eROSITA's sensitivity in the 0.2--8.0\,keV band, which is a factor ${\sim}$20--40 higher than \textit{ROSAT} \citep{2016A&A...588A.103B}, allows for
detection of sources that were too faint for \textit{ROSAT}'s soft
X-ray coverage or for hard X-ray observatories such as the INTErnational
Gamma-Ray Astrophysics Laboratory
\citep[\textit{INTEGRAL},][]{1993ESAJ...17..207W,2014A&A...567A...7D}
or the \textit{Neil Gehrels Swift} Burst Alert Telescope
\citep[\textit{Swift}/BAT,][]{2005SSRv..120..143B}.

In order to identify new X-ray binary candidates in the catalogue of the first
eROSITA all-sky survey (eRASS1), our team uses the typical X-ray and
optical properties (e.g., eROSITA fluxes and \textit{Gaia} magnitudes) of sources contained in the most recent
catalogues of heretofore known high-mass X-ray binaries \citep[HMXBs,][]{2023A&A...677A.134N} and
low-mass X-ray binaries \citep[LMXBs,][]{2023A&A...675A.199A} before performing
follow-up observations with other pointed instruments. Among the
sources flagged as X-ray binary candidates, our team obtained deeper
observations for four candidates by means of target of opportunity
(ToO) observations with the Nuclear Spectroscopic Telescope
Array~\citep[\textit{NuSTAR},][]{NuStar} and the X-ray Multi-Mirror
Mission~\citep[\textit{XMM-Newton},][]{XMM_mission}, with each observing two of the candidates. Analysis of one of the X-ray binary candidates followed-up with \textit{NuSTAR} has been presented by \citet{Doroshenko22}, while the two others have been reclassified as chromospherically active stars and will be discussed by Avakyan et al.\ (in prep.). 

\begin{figure}
    \centering
    \includegraphics[width=\linewidth]{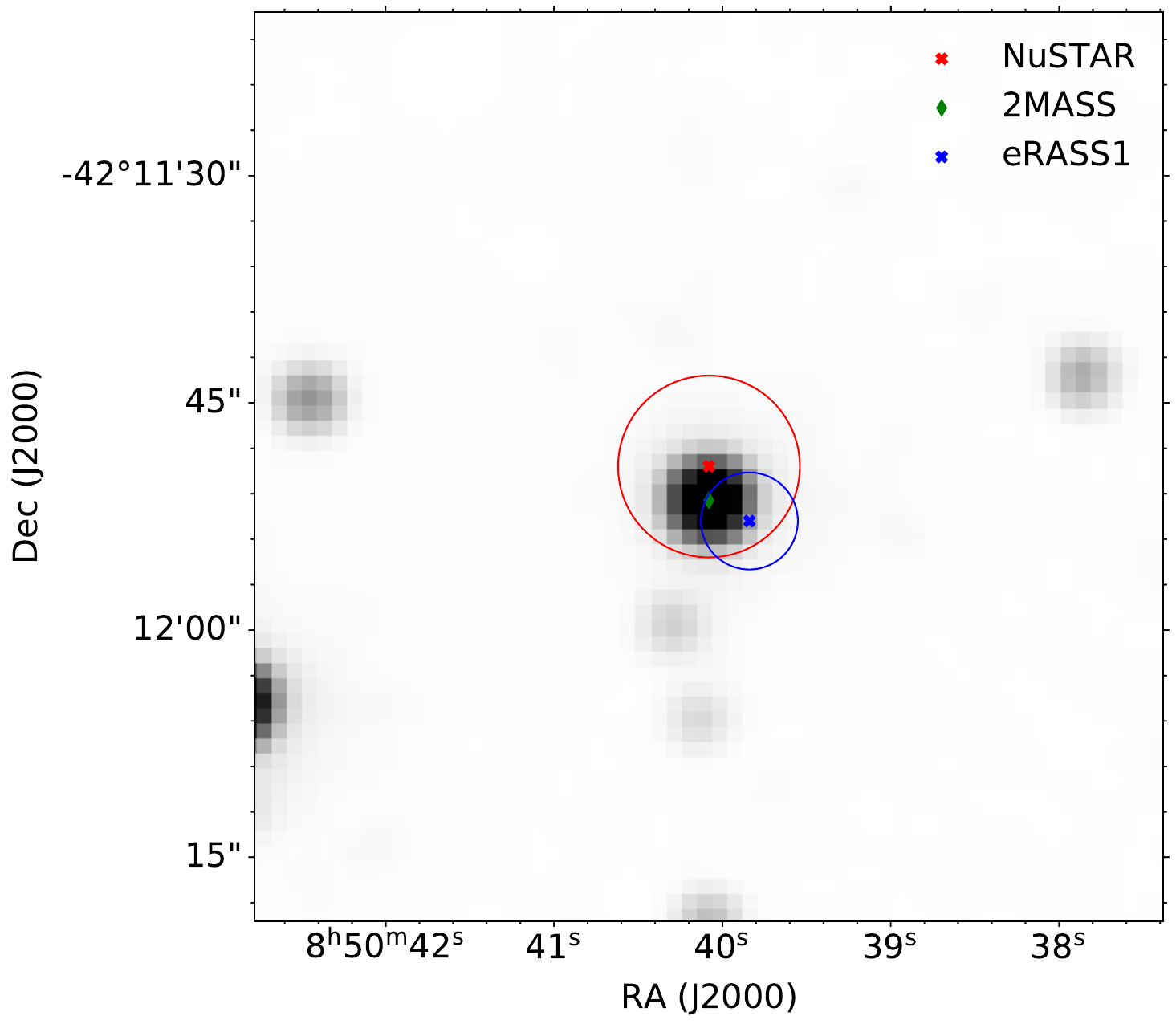}
    \caption{Infrared view of the optical counterpart from the Two Micron All Sky Survey \citep[\textit{2MASS},][]{2003tmc..book.....C}, with \textit{2MASS} \citep{2003tmc..book.....C}, \textit{NuSTAR} \citep{harrison2013} and \textit{eRASS1} \citep{Merloni23} error circles indicated in green, red, and blue respectively. The star is located $2\farcs{}7$ away from the eROSITA position of the source.} 
    \label{fig:opt_pos}
\end{figure}

This paper focuses on the analysis of eROSITA and \textit{NuSTAR}
observations of the fourth candidate, \fullname\footnote{The source name at the time of proposed follow-up was eRASS\,U\,J084850$-$420035, and has since been updated based on improved astrometry.} (hereafter \srcname), a
source that first came to our notice as an X-ray source found at
$\alpha_\mathrm{J2000.0} = 8^\mathrm{h}50^\mathrm{m}39\fs93$,
$\delta_\mathrm{J2000.0} = -42\degr11\arcmin57\farcs04$, with a
positional uncertainty of $1\farcs{}61$ \citep{Merloni23}, during the first eROSITA All-Sky Survey (eRASS1). The system had been previously detected in X-rays (as Swift J0850.8$-$4219) in
the 105\,month \textit{Swift}/BAT All Sky Hard X-ray Survey by the
\textit{Swift}/BAT-instrument on board the Neil Gehrels Swift Observatory
\citep{SWIFT} at the position
$\alpha_\mathrm{J2000.0} = 8^\mathrm{h}50^\mathrm{m}39\fs80$,
$\delta_\mathrm{J2000.0} = -42\degr11\arcmin51\farcs90$, where it was
flagged as an unidentified source \citep{2018ApJS..235....4O}. We
flagged \srcname as an X-ray binary candidate due to its positional
coincidence with the star UCAC2 13726137 (Fig.~\ref{fig:opt_pos}),
which is only $3''$ away, and has since been identified as a K-type supergiant \citep[][hereafter, \citetalias{2024MNRAS.528L..38D}]{2024MNRAS.528L..38D}. This star is contained in the \textit{Gaia}
catalogues and is reported to have a G-band magnitude of 13.354$\pm$0.002\,mag, and a parallax of
$0.0827\pm0.0139$\,mas \citep{2016A&A...595A...1G, 2023A&A...674A...1G}, corresponding to a
geometric distance of $7.45^{+0.75}_{-0.71}$\,kpc
\citep{2021AJ....161..147B}. 

Here, we report on the analysis of the eROSITA
data, a pointed observation with \textit{NuSTAR}, and archival observations from the ESO Very Large Telescope (VLT)'s spectrograph \xsh \citep{2011A&A...536A.105V} of the optical counterpart, with the
primary focus on determining the nature of its compact object. We compare our results with the findings of \citetalias{2024MNRAS.528L..38D}, who studied Swift/\textit{XRT} data and a near-infrared (NIR) spectrum obtained using the Southern Astrophysical Research Telescope (SOAR)'s TripleSpec spectrograph \citep{2014SPIE.9147E..2HS}. The
paper is structured as follows. In Sect.~\ref{section:data}, we
describe the observations obtained and analysed, as well as report on information gathered from several survey instruments. This is followed by
inferences on the behaviour of the optical counterpart in
Sect.~\ref{section:optical}. Sect.~\ref{section:x-ray_analysis}
discusses the timing and spectral analysis of the X-ray
data, based on which we discuss likely physical scenarios for the
system in Sect.~\ref{section:discussion}, before concluding on our findings in Sect.~\ref{section:conclusion}.

\section{Observations and data reduction}\label{section:data}

\subsection{eROSITA}
\srcname has been observed by eROSITA four times, once during
each complete All-Sky Survey conducted thus far (Table
\ref{tab:eRO_data}). The data were obtained within the framework of
the \textit{eROSITA\_DE} consortium. We extracted the data products
using \texttt{evtool} and \texttt{srctool} from the eROSITA
data analysis software \citep[eSASS,][]{2022A&A...661A...1B}, version 211214, processing version c020,
and the High Energy Astrophysics Software (HEASOFT) version 6.29. eSASS computed source and background extraction regions for this source to be $20''$ and an annulus with inner and outer radii ${\sim}48''$ and ${\sim}250''$, respectively. 
\begin{table}
  \centering
  \caption{Observing log of the four eROSITA snapshots of \srcname and the follow-up observation obtained by \textit{NuSTAR}.}
  \renewcommand{\arraystretch}{1.4} 
  \renewcommand{\tabcolsep}{1.2mm}
  \begin{tabular}{cccc}
    \hline \hline
    eROSITA Scan & MJD & Exposure [s] & Counts \\
    \hline
    1 & 58989.057 & 317 & 88 \\
    2 & 59172.359 & 210 & 40 \\
    3 & 59354.932 & 209 & 36 \\
    4 & 59539.401 & 237 & 51 \\
    \hline
    \textit{NuSTAR} Pointing & MJD & Exposure [s] & Counts \\
    \hline
    FPMA & 59347.196 & 55638 & 15277 \\
    FPMB & 59347.196 & 55183 & 15158 \\
    \hline
  \end{tabular}
  \tablefoot{The table lists the mid-time of each observation together with net exposures and total counts (including background).}
  \label{tab:eRO_data}
\end{table}

\begin{figure}
  \centering
    \includegraphics[width=\linewidth]{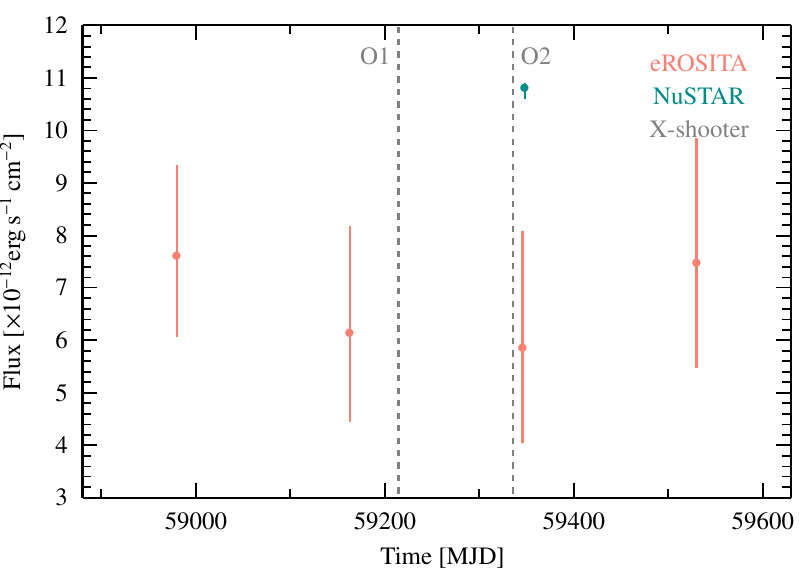}
    \caption{The 3--10\,keV X-ray flux from each observation of \srcname as a functin of time, corrected for absorption. The eROSITA fluxes are displayed at the start time of each scan. The third eROSITA observation was only four days prior to the pointed \textit{NuSTAR} observation. The \textit{NuSTAR} flux is brighter by a factor ${\sim}$2, which might indicate slight source-intrinsic variability, even when taking into account eROSITA's known calibration uncertainties \citep{2024A&A...688A.107M}. The observation times of the optical observations from \xsh are marked by gray dashed lines  as O1 and O2.} 
    \label{fig:lum-states}
\end{figure}

Since eROSITA performs a slew survey, the total exposure of sources
during each All-Sky Survey is only a few 100\,s, taken during several
$\sim$40\,s long passes of the source through eROSITA's field
of view, which are separated by 4\,h due to the rotation of \textit{SRG}. 

\subsection{\textit{NuSTAR}}
Due to the short exposure times of the eROSITA slew survey, with the
exception of for the brightest X-ray sources, follow-up observations with other instruments are necessary to accumulate data that are suited
for a detailed quantitative analysis that would allow us to deduce physical information of the source. We therefore triggered
follow-up observations of \srcname with \textit{NuSTAR},
launched on 2012 June 13 into an orbit with a period of about
97\,minutes. It comprises two detector arrays, focal plane modules A and B (FPMA and FPMB), which operate in the 3--79\,keV energy range \citep{harrison2013}. 

\textit{NuSTAR} observed \srcname between MJD~59347.187 and 59348.388 (OBSID: 80660302002), close to the third
eROSITA observation (Fig.~\ref{fig:lum-states}) with an
effective exposure time of 55.6\,ks. We carried out the data reduction
and processing using \textit{NuSTAR's} analysis software NuSTARDAS,
version~2.1.1, v20211115, of the \textit{NuSTAR} calibration database (CALDB),
and HEASOFT v6.29. We reduced the raw data to produce cleaned event
lists using \texttt{nupipeline} and applied \texttt{nuproducts} to
extract spectra and barycentered lightcurves from circular regions of
${\sim}45''$ radius separately for FPMA and FPMB. The background
regions were circles with ${\sim}117''$ radius, offset from the source
but on the same detector chip.

\subsection{Other X-ray data}
In addition to the \textit{NuSTAR} follow-up, we searched for the source in other X-ray survey data. The results are listed in Table~\ref{tab:survey_data} and described as follows. We refer to the 157-Month lightcurve provided by the 157-Month \textit{Swift}/BAT All Sky Hard X-ray Survey (Lien et al., in prep.), which spans the period from the detection of the source in 2004 December until 2017 January. We obtained Crab-weighted lightcurves of \srcname's \textit{Swift}/BAT counterpart, Swift J0850.8$-$4219, from the \textit{Swift}/BAT Hard X-ray Survey webpage hosted by the High Energy Astrophysics Software ARChive (HEASARC) server\footnote{\url{https://swift.gsfc.nasa.gov/results/bs157mon/}}, and determine the flux as listed there. Archival \textit{Swift}/XRT observations with serendipitous detection of the source were obtained from HEASARC. We also estimated fluxes, and upper limits where fluxes were unattainable, as observed by other survey missions at the source position of \srcname, namely by \textit{ROSAT}, \textit{INTEGRAL}, and the Mikhail Pavlinsky Astronomical Roentgen Telescope-X-ray Concentrator \citep[\textit{ART-XC},][also on board SRG]{2021A&A...650A..42P}. 

The upper limits on \textit{ROSAT} count rate were obtained using the upper limit server HIgh-energy LIght-curve GeneraTor\footnote{\url{http://xmmuls.esac.esa.int/upperlimitserver/}} \citep[HILIGT,][]{2022A&C....3800531S, 2022A&C....3800529K}. We converted them to a flux upper limit with WebPIMMS\footnote{\url{https://heasarc.gsfc.nasa.gov/cgi-bin/Tools/w3pimms/w3pimms.pl}}
assuming a powerlaw with an index of $\Gamma=0.5$, and an $N_\mathrm{H}=1.5\times10^{22}\,\mathrm{cm}^{-2}$ (see Sect.~\ref{subsec:spec_analysis}). 

We determined the \textit{INTEGRAL} flux upper limits in the 30--50\,keV band from merged IBIS/ISGRI \citep{2003A&A...411L.131U, 2003A&A...411L.141L} spectra of all IBIS Science Windows during which the source position had an off-axis angle of ${<}10^{\circ}$. We used version 11.2 of the Off-line Scientific Analysis (OSA)\footnote{\url{https://www.isdc.unige.ch/integral/download/osa/doc/11.2/osa_um_ibis/man_html.html}}  for the data extraction. The resulting spectrum includes a total of $2.8\,\mathrm{Ms}$ of exposure time. We described its shape with a simple powerlaw model to determine a flux upper limit.  

The ART-XC flux for \srcname's counterpart in the 4--12\,keV band was obtained from the recently published catalog of all sources detected by \textit{SRG}/ART-XC in its first year \citep{2022A&A...661A..38P}, where it has been independently identified as a potential HMXB candidate. 

\textit{Swift}/XRT fluxes were obtained by fitting an absorbed powerlaw to the spectra extracted at the source position, with extraction regions spanning ${\sim}25''$ in radii. The fitting resulted in a relatively unconstrained photon index of $\Gamma=0.6^{+1.8}_{-0.9}$ with $N_\mathrm{H}=7^{+12}_{-8}\times10^{22}\,\mathrm{cm}^{-2}$. We do not use these spectra for further spectral analysis, in favour of the eROSITA data which provide more overall counts. 

\begin{table}
  \centering
  \caption{\srcname as seen by different observatories, over the years.}
  \renewcommand{\arraystretch}{1.4} 
  \renewcommand{\tabcolsep}{1.2mm}
  \begin{tabular}{cccc}
    \hline \hline
    Observatory & Time & Energy Range & Flux \\
    & [year] & [keV] & $\times 10^{-12}\,\mathrm{erg}\,\mathrm{s}^{-1}\,\mathrm{cm}^{-2}$\\
    \hline
    \textit{ROSAT} & 1990 & 0.2--2.0 & ${<}$1.48\\
    \textit{SRG}/eROSITA$^{*}$ & 2020--2021 & 0.2--2.0 & $1.04^{+0.46}_{-0.31}$ \\  
    \textit{Swift}/XRT & 2011 & 0.5--10.0 & 5.01$_{-1.7}^{+0.74}$\\
    & 2011 & 0.5--10.0 & 6.53$_{-1.41}^{+1.40}$ \\
    \textit{SRG}/eROSITA$^{*}$ & 2020--2021 & 0.5--10.0 & $7.24^{+2.30}_{-1.99}$ \\   
    \textit{SRG}/ART-XC & 2020 & 4.0--12.0 & 8.8$_{-2.6}^{+3.2}$\\
    \textit{NuSTAR} & 2020 & 4.0--12.0 & 10.6$\pm0.1$\\
    \textit{INTEGRAL}$^{*}$ & 2003--2023 & 30.0--50.0 & ${<}$4.8 \\
    \textit{NuSTAR} & 2020 & 30.0--50.0 & 4.4$\pm0.5$\\
    \textit{Swift}/BAT$^{*}$ & 2004--2017 & 14.0--195.0 & 12.08$_{-1.02}^{+2.05}$ \\
    \textit{NuSTAR} & 2020 & 14.0--55.0 & 19.1$\pm$0.2\\
    \hline
  \end{tabular}
  \tablefoot{$^{*}$The reported flux is the averaged flux over multiple snapshots during the specified time range.}
  \label{tab:survey_data}
\end{table}

\subsection{\xsh}
We also used data from the two observations of the potential companion
star of \srcname in the optical and near-infrared found in the ESO archive. These data were
obtained using the VLT's spectrograph \xsh
\citep{2011A&A...536A.105V}. \xsh is a medium resolution
spectrograph that has three spectroscopic arms, operating in ultraviolet-blue (UVB; 300--559.5\,nm), visible
(VIS; 559.5--1024\,nm), and near-infrared (NIR; 1024--2480\,nm). The
observations were carried out four months apart, on
2021 January 11 (MJD 59225) and 2021 May 14 (MJD 59348). Each observation had an exposure of ${\sim}$1000\,s
in each of the three bands covered by \xsh. 

Finally, the counterpart is also monitored by the All Sky Automated Survey for Supernovae \citep[ASAS-SN,][]{2017AcASn..58...41S}, where we used the lightcurve as hosted on the ASAS-SN Lightcurve Server\footnote{\url{https://asas-sn.osu.edu}}. 

\subsection{Search for counterparts in other wavelengths}
Apart from the optical and X-ray regimes, X-ray binaries can also sometimes show up in other wavelengths, as microquasars \citep{2019ApJ...871...26K} or accreting millisecond X-ray pulsars (AMXPs) in radio \citep[e.g.][]{2020MNRAS.492.1091G}, or as $\gamma$-ray binaries emitting $\gamma$-rays \citep{2020mbhe.confE..45C}. Emission, or lack thereof, in these wavelengths can serve as yet another distinguishing feature that allows us to place the system among the various subclasses of X-ray binaries, and hint at the nature of the compact object.   

With this in mind, we performed a search for counterparts at the coordinates of \srcname with survey data from the Fermi Gamma-ray Space Telescope's Large Area Telescope \citep[\textit{Fermi/LAT},][]{2009ApJ...697.1071A}. The search returned no reliable counterparts within $25'$ of the source position. In order to look for radio counterparts, we queried HEASARC's regularly updated Master Radio Catalog\footnote{\url{https://heasarc.gsfc.nasa.gov/w3browse/master-catalog/radio.html}} which  provides access to all available radio source catalogs. We did not find any reliable radio counterparts within the error circle of the eROSITA position of \srcname or within that of the optical counterpart. 

\section{The nature of the optical counterpart UCAC2 13726137}
\label{section:optical}

\subsection{Narrowing down the spectral type and luminosity class}
\label{subsec:sp_type}
We start our discussion of the nature of \srcname with a look at the
optical counterpart, since this can help us narrow down the possible configurations X-ray binaries may exist in, which are typically classified based on the mass and spectral type of the donor star \citep{2017AcASn..58...41S,2023arXiv230802645F}.
HMXBs are typically subdivided into Be X-ray Binaries (BeXRBs) and Supergiant X-ray Binaries (SgXBs), with the companion stars being non-supergiant Be-type stars and massive supergiant stars, typically of spectral type OB, respectively \citep{2011Ap&SS.332....1R, 2023arXiv230802645F}. In rare cases, the high mass donor can also be a late-type red supergiant (RSG), as in 4U~1954+31 \citep{hinkle_m_2020}. LMXBs are instead characterised by late-type dwarf companions \citep[see, e.g.,][for a review]{bahramian.book..120B},
or, in a smaller subset of cases, M-giant donors. This subset of sources is called Symbiotic X-ray Binaries \citep[SyXBs,][]{bozzo_symbiotic_2022,yungelson_wind-accreting_2019}. They are supposed to host highly magnetised neutron stars accreting from the slow stellar wind of their M~giant optical counterparts.  

Optical spectra enable us to discern between these possible donor types, based on the presence of various emission lines and overall spectral shape \citep[see, e.g.,][]{2012A&A...538A.123M}. Be-type stars, typically defined as non-supergiant stars, show Balmer emission lines in their spectra at any time \citep{1988PASP..100..770S}, which are attributed to the presence of a decretion disk. OB-type supergiants can also exhibit some Balmer lines in emission, albeit shallower, but typically have several metal lines in emission \citep{2012MNRAS.423..284A, 2019ApJS..241...32L}. Late-type giant and supergiant stars typically do not show strong Balmer lines in emission, and are instead characterised by the presence of molecular absorption bands, mainly TiO in the optical and CO in the NIR regimes \citep{serote_roos_spectra_1996}. M giants in SyXBs do sometimes also show key lines in emission such as H$\alpha$ \citep{davidsen_optical_1977}. It should also be noted that M giants in accreting systems have been predominantly studied in the context of Symbiotic stars (SySts) hosting white dwarfs as the compact object \citep{merc_new_2019}. 

\begin{figure*}
  \centering
  \includegraphics[width=\textwidth]{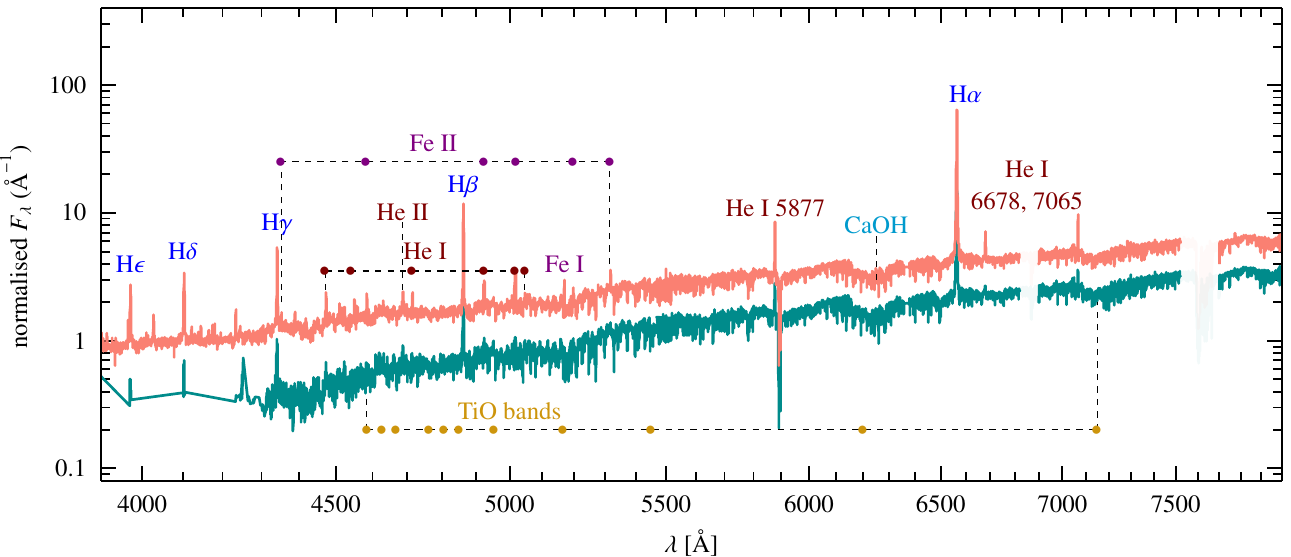}
  \vspace{-0.3cm}
  \includegraphics[width=\textwidth]{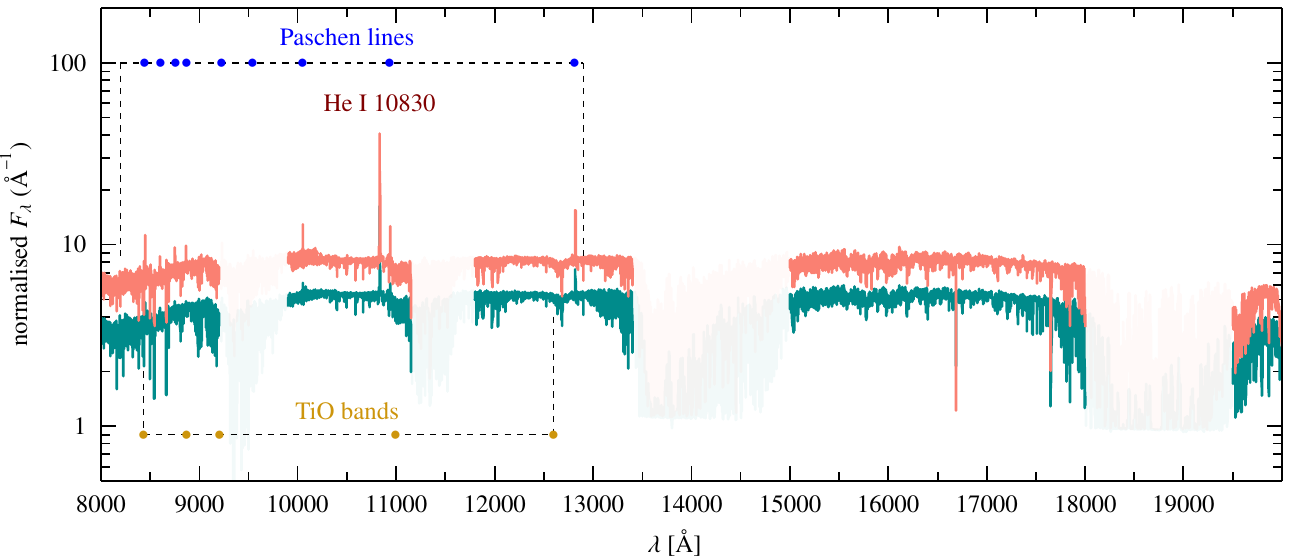}
  \vspace{-0.3cm}
  \includegraphics[width=\textwidth]{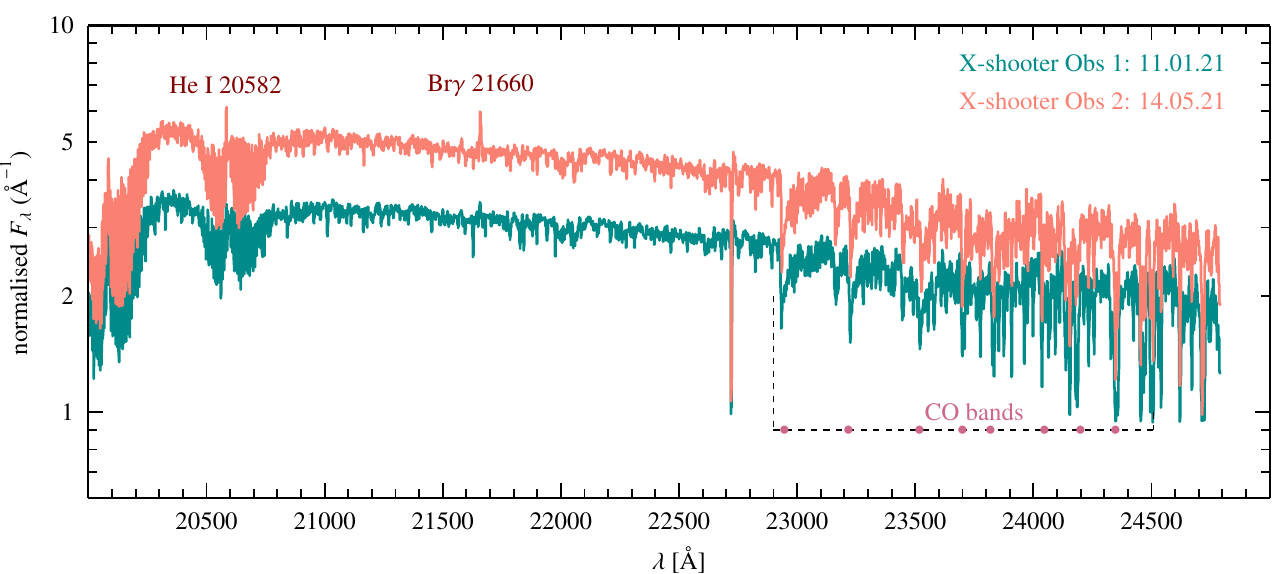}  
  \caption{Dereddened \xsh spectra of the optical counterpart.
     Observation~2 shows an increase
     in flux with stronger emission lines. Prominent features include the Balmer series, several \ion{He}{i}  and \ion{Fe}{ii} lines, and TiO and CO absorption bands. Telluric regions are greyed out.}
    \label{fig:opt_spec}
\end{figure*}

Although \citetalias{2024MNRAS.528L..38D} already characterised the optical counterpart of \srcname as a K4--5 supergiant, we nevertheless conduct an independent study, including a wider wavelength regime using the \xsh data. The \xsh spectra shown in Fig.~\ref{fig:opt_spec} clearly
indicate several lines in emission, most notably the Balmer lines. These are strong indicators for the presence of a hot circumstellar
decretion disk around a Be-type star \citep{1988PASP..100..770S}. However, the spectra also show prominent TiO bands at 6250\,\AA, 7150\,\AA, and 8500\,\AA, and very evident CO bands in the NIR spectra, strongly supporting the late-type characterisation invoked by \citetalias{2024MNRAS.528L..38D}. 

In order to determine the spectral type, we employ methods to discern spectral type in SyXBs as put forth by \citet{kenyon_cool_1987}, who established that the depth of TiO bands acts as an indicator for spectral type and temperature of the star, independent of the effects of reddening, and have been similarly applied elsewhere \citep[e.g.,][]{munari_infrared_2018}. We determine the [TiO]$_{1}$ and [TiO]$_{2}$, and the [Na] index, as defined by \citet[][their equations 1, 2 and~4]{kenyon_cool_1987}, and use them to compute the spectral type according to equations (5) and (6) of the same paper, 
\begin{equation}
\mathrm{ST}_{1} = 1.75 + 9.31[\mathrm{TiO}]_{1}
\label{eq:st1}
\end{equation}
and 
\begin{equation}
  \mathrm{ST}_{2} = 1.83 + 10.37[\mathrm{TiO}]_{2} - 3.28[\mathrm{TiO}]_{2}^2,
  \label{eq:st2}
\end{equation} 
where the spectral type, ST is $-6$ for K0 stars, $0$ for M0 stars, and $+6$ for M6 stars. 

The indices determined from the \xsh spectra are listed in Table~\ref{tab:spectral_indices}. Using Eqs.~\ref{eq:st1} and~\ref{eq:st2}, we find $\mathrm{ST}_{1}$ ranging from ${\sim}2.86$--3.14 and $\mathrm{ST}_{2}$ ranging from ${\sim}1.80$--1.98 between the two observations, pointing to a spectral type closer to ${\sim}$M2--3, as opposed to the K4--5 characterisation invoked by \citetalias{2024MNRAS.528L..38D}. We compared UVES \citep{bagnulo_uves_2003} spectra corresponding to an M~type and K~type supergiant with the \xsh data, and find the TiO bands observed in UCAC2 13726137 are only visible in the M~type spectrum. The conclusion that UCAC2 13726137 is a M~type star is further supported by the correlation between the equivalent width of the $\mathrm{CO}(2,0)$ band and spectral type as put forth by \citet{davies_massive_2007}, where our estimated values populate the top right of their Fig.~2, already hinting at a luminosity class of~I. 

\begin{figure}
  \centering
  \includegraphics[width=\linewidth]{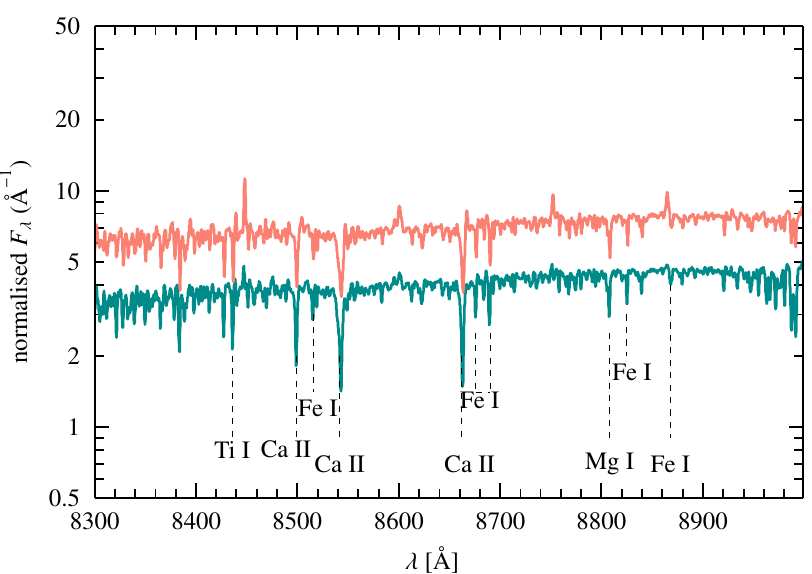}  
  \caption{The 8300--9000\,\AA\xspace region of the NIR spectrum. Several key absorption lines used for classification of the donor star are indicated here.}
    \label{fig:cenarro_region}
\end{figure}

We nevertheless also use other diagnostics to estimate the luminosity class. We use the equivalent width (EW) of the line blend constituting \ion{Ti}{i}, \ion{Fe}{i} and CN around 8468\,\AA\xspace (see Table~\ref{tab:eqwidths}), which sits in the parameter space occupied by the handful of supergiants considered by \citet{negueruela_massive_2011} in their Fig.~5, where they correlate EW(8468\AA) with spectral type. \citet{messineo_red_2017} analysed a sample of red supergiants, and find that they lie in the bottom right while correlating the equivalent widths of the \ion{Mg}{i} line and the $\mathrm{CO}(2,0)$ band. We compute the equivalent width of the $\mathrm{CO}(2,0)$ band as defined by \citet{messineo_new_2021} for both observations. Some of the lines in this spectral region are shown in Fig.~\ref{fig:cenarro_region}. We find our results for these lines consistent with the positions occupied by supergiants in this parameter space, compared to giants. We also estimated the equivalent widths and indices proposed by \citet{dorda_characterisation_2016}, but found our overall values systematically larger, likely, as they indicate, due to differences between red supergiant properties between the Magellanic Clouds -- where their sample lies -- and the Galaxy. We therefore do not use their characterisation, but tentatively suggest that the star is of luminosity class~I, in agreement with \citetalias{2024MNRAS.528L..38D}. Below, we retrace the steps taken by \citetalias{2024MNRAS.528L..38D} and compare our results. 

\begin{table}
\centering
  \caption{The spectral indices used in this work.}
  \renewcommand{\arraystretch}{1.4} 
  \renewcommand{\tabcolsep}{1.2mm}
  \begin{tabular}{ccc}
    \hline \hline
    Index & Observation 1 & Observation 2 \\
    \hline
    $[\mathrm{TiO}]_{1}$ & 0.12 & 0.15 \\
    $[\mathrm{TiO}]_{2}$ & -0.002 & 0.015 \\
    $[\mathrm{Na}]$ & 0.12 & 0.07 \\
    \hline
  \end{tabular}
  \tablefoot{We refer to \citet{kenyon_cool_1987} for the definition of these indices.}
  \label{tab:spectral_indices}
\end{table}

\begin{figure}
  \centering
  \includegraphics[width=\linewidth]{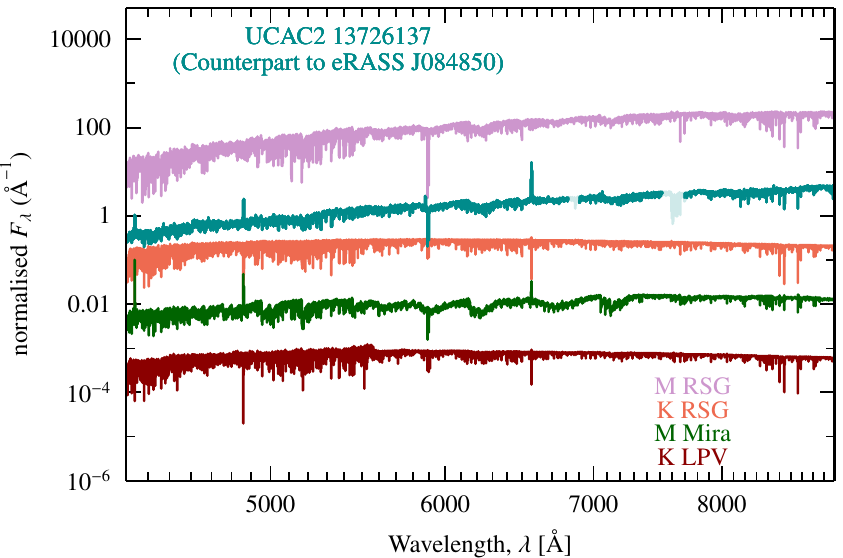}  
  \caption{Comparison of the dereddened \xsh spectra of the optical counterpart to \srcname with spectra of a variety of late type stars, obtained from the \xsh spectral library. Although the K-type supergiant has some absorption lines in common with that of UCAC2 13726137, the latter's overall shape seems to agree better with the M-type supergiant, in addition to the presence of TiO bands. Of special intrigue is the remarkable similarity in spectral shape and presence of the same emission lines in the Mira variable.}
  \label{fig:comp_opt_specs}
\end{figure}

\begin{table}
    \centering
    \caption{Equivalent widths of spectral lines obtained from the \xsh spectra, as well as other diagnostics.}
    \renewcommand{\arraystretch}{1.4} 
    \renewcommand{\tabcolsep}{1.2mm}
    \begin{tabular}{cccc}
    \hline
    \hline
        Line & Wavelength & Obs.~1 & Obs.~2 \\
       & \AA & \AA & \AA \\
       \hline
         H$\alpha$ & 6563 & 16.97 & 23.11 \\
         \ion{Ca}{ii} & 8498 & 9.35 & 7.78 \\
         \ion{Fe}{i} & 8514.1 & 3.34 & 2.50 \\
         \ion{Ti}{i} & 8518.1 & 2.6 & 1.7  \\ 
         \ion{Mg}{i} & 17111 & 1.53 & 1.36 \\
         \ion{Na}{i} & 22050 & 5.59 & 5.13 \\
         \ion{Ca}{i} & 22063 & 0.53 & 0.72 \\ 
         CO(2,0) & 23000 & 38.11 & 30.70 \\ 
         \hline
         & \multicolumn{2}{c}{Other diagnostics and indices} & \\
         \hline
         \multicolumn{2}{c}{$\log(\mathrm{EW}(\mathrm{CO})/(\mathrm{EW}(\mathrm{Na})+\mathrm{EW}(\mathrm{Ca})))$} & 0.79 & 0.719\\
         \multicolumn{2}{c}{J8} & 0.26 & 0.25 \\
         \multicolumn{2}{c}{J9} & 1.36 & 1.27\\
         \multicolumn{2}{c}{J10} & 4.83 & 3.71 \\
         \multicolumn{2}{c}{\ion{Ti}{i}, \ion{Fe}{i} and $\mathrm{CN}$ line blend} & 3.52 & 1.80 \\
         \hline
    \end{tabular}
    \tablefoot{See text for a discussion.}
    \label{tab:eqwidths}
\end{table}

\subsection{Retracing the steps undertaken by De et al. 2024}

\citetalias{2024MNRAS.528L..38D} show a comparison of their SOAR~T-Spec spectrum of UCAC2 13726137 to NIR spectra of an M~giant and a K~supergiant, eventually concluding that a K~supergiant is the better fit. In this section we take a deeper look at the diagnostics invoked by them, and apply them to the \xsh spectra. 

We find our measured values for the equivalent width of the $\mathrm{CO}(2,0)$ band comparable to that determined by \citetalias{2024MNRAS.528L..38D}. Further, we measure equivalent widths of the absorption features at \ion{Na}{i} and \ion{Ca}{i}, although we find them to be more complex in the \xsh spectra than in the T-Spec data. We also confirm that $\log(\mathrm{EW}(\mathrm{CO})/(\mathrm{EW}(\mathrm{Na})+\mathrm{EW}(\mathrm{Ca})))$ lies in the parameter space for evolved late-type stars in the diagnostic put forth by \citep{ramirez_luminosity_1997}. The respective values are tabulated in Table~\ref{tab:eqwidths}. 

In accordance with \citetalias{2024MNRAS.528L..38D}, we find CN absorption features in the Y + J band, but in addition, also find a TiO band at 11000\,\AA. On closer inspection, the SOAR Tspec spectrum studied by \citetalias{2024MNRAS.528L..38D} also reveals the presence of the TiO absorption band, although it is masked by the strong \ion{He}{i} emission line at 10830\,\AA. As mentioned earlier, the presence of TiO bands throughout the optical and infrared regimes points to a later spectral type than K~4--5. \citet{dorda_spectral_2016} suggest that deriving the spectral type from TiO bands results in underestimating the temperatures and suggest using atomic line features such as those from \ion{Ti}{i}  and \ion{Fe}{i} instead. However, as previously stated by \citet{dicenzo_atomic_2019}, their results are based on a sample of stars ranging between G--M3 in spectral type, with the M stars making up a very small fraction. \citet{dicenzo_atomic_2019} reiterate on the reliability of using the TiO band for late spectral types to determine the effective temperature. The M-type description of UCAC2 13726137 is further supported by correlations between effective temperature and equivalent widths of \ion{Na}{i}, \ion{Ca}{i}, and $\mathrm{CO}(2,0)$ lines, respectively, \citep[see Fig.~9 in][]{ramirez_luminosity_1997}, where our values all correspond to a cool star with temperatures between 3200--3600\,K. However, we are in complete agreement with \citetalias{2024MNRAS.528L..38D} on the luminosity class of the star, based on the diagnostics used by them employing the equivalent widths of lines from \ion{Mg}{i} and $\mathrm{CO}(2,0)$, and the J8, J9, and J10 indices from \citet{messineo_new_2021}. We therefore tentatively reclassify UCAC2 13726137 as an M2--3 supergiant. 

\subsection{Disentangling the emission lines}
As alluded to in Sect.~\ref{subsec:sp_type}, emission lines are not typically observed in late-type supergiants. The emission lines must arise from matter surrounding the star, be this intrinsic to the star, or owing to the binary configuration of the system. However, if the emission lines are due to photo-ionisation by the compact companion, as suggested by \citetalias{2024MNRAS.528L..38D} , the transient nature of some of them observed in the optical spectrum should correlate with variability in the X-rays. An example of variability in the emission lines between the two observations can be seen in the case of H$\alpha$, shown in Fig.~\ref{fig:halpha_line}. This is hard to reconcile with the fairly persistent X-ray flux of \srcname as detected by eROSITA in each scan. Other lines such as those from \ion{He}{i} and \ion{He}{ii}, also depicted in Fig.~\ref{fig:halpha_line}, are variable but markedly less so in comparison. 

An obvious analogue to these spectra are the optical spectra of SySts, which are characterised by having the cool spectral continuum of a K--M giant, with several hydrogen, helium, and iron lines in emission, typically also of high ionisation states such as [\ion{Fe}{vii}] \citep{kenyon_nature_1984}. Although we find several hydrogen and helium lines in the \xsh spectra, we do not find lines such as [\ion{Fe}{VII}$_{\mathrm{\lambda},6087}$] or [\ion{He}{ii}$_{\mathrm{\lambda},10123}$]. Due to this, in conjunction with the apparent lack of pronounced X-ray variability, we attempted to find other optical spectra showing similar lines in emission. We start with a comparison of our spectra to \xsh spectra of late-type giants and supergiants. In particular, we identified KM-type giants and supergiants in the \xsh spectral library (XSL) published as part of the third \xsh data release \citep{verro_x-shooter_2022}. The giants in the sample are typically variable stars -- Mira variables, or long period variables (LPV), which are both known to show emission lines in their spectra. Some of these are shown in Fig.~\ref{fig:comp_opt_specs}. 

We find that the red supergiants in the sample show similar lines in absorption with the strength of the TiO bands increasing with later spectral types, but there was no instance of emission lines as observed in UCAC2 13726137. On the other hand, we noticed a remarkable consistency with the spectra of Mira variables which display several of the same lines in emission as seen in one observation of UCAC2 13726137. 

\begin{figure}
  \centering
  \includegraphics[width=\linewidth]{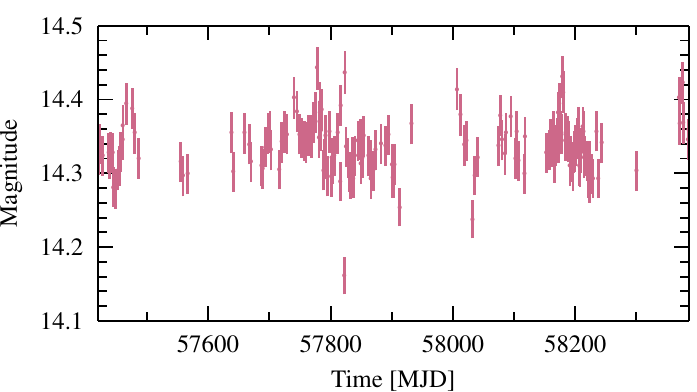}  
  \caption{ASAS-SN lightcurve of \srcname. The lightcurve shows semiperiodic variable behaviour. }
    \label{fig:asassn_lc}
\end{figure}

\begin{figure}
  \centering
  \includegraphics[width=\columnwidth]{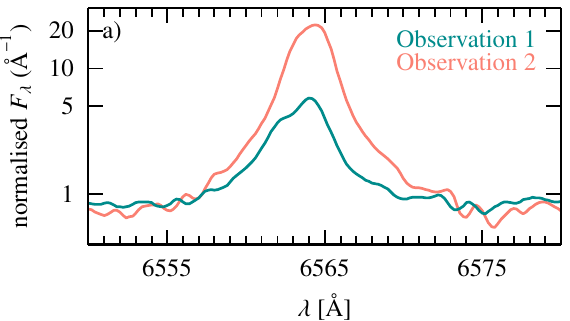}
  \includegraphics[width=\columnwidth]{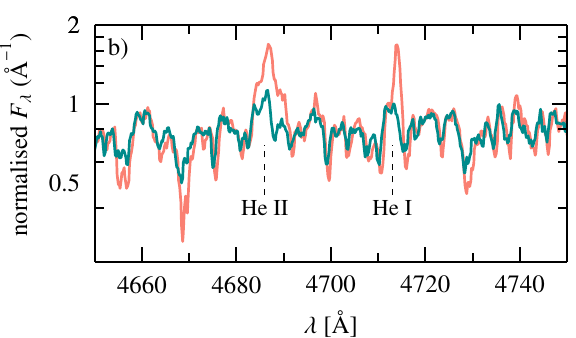}
  \caption{\textbf{a} H$\alpha$ line as seen in both \xsh spectra, normalised for comparison. The second observation shows a stronger and single-peaked line profile, compared to an asymmetric line profile in the first data set. The asymmetry is likely due to the contribution of \ion{He}{ii} $\lambda6560$, also being in emission. \textbf{b} \ion{He}{ii} $\lambda4686$ and \ion{He}{i} $\lambda4713$ are in emission, in both \xsh spectra but do not show as much variability. We normalised the two spectral continua relative to each other, to compare the behaviour of the emission lines.}
    \label{fig:halpha_line}
\end{figure}

The similarity of the \xsh spectra with that of Mira variables deserves a deeper look. Mira variables are cool giants of M-type which show regular variability, with periodicities of more than a few 100\,days, and amplitudes $\geq$2.5\,mag \citep{percy_understanding_2007}. In order to address the possibility of \srcname's optical companion indeed being a Mira variable, we looked for evidence of variability in the optical regime. Based on the ASAS-SN lightcurve (Johnson V-band, $\sim$520\,nm), UCAC2 13726137 has been reported to show semi-periodic photometric variability with a periodicity of ${\sim}$39.24\,d \citep{2021MNRAS.503..200J}. The lightcurve is displayed in Fig.~\ref{fig:asassn_lc}, confirming semiregular variability with an amplitude of about $\pm0.1$\,mag. According to \citet{2021MNRAS.503..200J}, it was therefore classified as a semiregular variable star (SRV). These objects are late spectral type giants or supergiants showing variability \citep{samus_general_1997,percy_understanding_2007}. They differ from Mira variables in their lower variability amplitude and less regular periodic behaviour. It is unclear, however, what this periodicity exactly pertains to in the case of UCAC2 13726137. With Mira variables, the periodic change in brightness is attributed to the expansion and contraction of these M giants, which is also responsible for the presence of the emission lines. A similar scenario can be invoked for SRVs \citep{moon_combining_2008,kiss_multiperiodicity_1999}. 

Although epoch folding \citep{1983ApJ...272..256L} on the publically available ASAS-SN data does not yield a significant periodicity, and we therefore do not use it to get any orbital constraints, the ASAS-SN lightcurve serves to confirm the variable nature of the star. Semiregular variables are generally classified into four subclasses based on the kind of variability observed: SRa, SRb, SRc, and SRd, with their typical features listed in Table~\ref{tab:SR_class}, based on the General Catalogue of Variable Stars \citep[GCVS,][]{samus_general_1997,samus_general_2017}, confirming that there are indeed supergiants that are classed as `semi-regular'. Although we do not find many precedents for M~supergiant emission line spectra, they are not unheard of, as can be seen in the case of the unusually late-type red supergiant WOH G64 discussed by \citet{levesque_physical_2009}. The emission lines in this object are attributed to the circumstellar environment, with the suggested dominant source of ionisation being shock heating due to pulsations, both common characteristics of M supergiants \citep{skinner_circumstellar_1988}, not unlike the variable stars discussed above. Moreover, as \citet{skinner_circumstellar_1988} reiterate, the lack of significant variability in the optical regime is insufficient evidence for the absence of pulsations, since red supergiants peak in the infrared, where we have sparse monitoring data. 

The alternative scenario for WOH G64 explored by \citet{levesque_physical_2009} is a binary nature of the system, which still needs probing into in our case. As briefly mentioned above, SySts and SyXBs are observed to have cool spectra with emission lines due to the hot companion. We therefore explore the emission lines that are typically observed in this case. Early studies of symbiotic binaries \citep{proga_he_1994} report several \ion{He}{i} lines in emission, although at the time the ionising region was simply considered to be a hot nebula  and not an accreting compact object. Balmer lines are also seen in emission in Miras \citep{fox_shock_1984} and several SySts \citep{miszalski_symbiotic_2013}, and accreting X-ray binaries show at least H$\alpha$ in emission \citep[e.g.,][]{torres_vlt_2015,bassa_optical_2009}. In the latter case, they are often taken as an indicator for the presence of an accretion disk \citep[][and references therein]{fender_anticorrelation_2009}. A prototypical SyXB hosting a neutron star is GX\,1$+$4, which has a rich emission line spectrum \citep{davidsen_optical_1977} attributed to a persistent accretion disk, but the \ion{Fe}{ii} line forest around its H$\alpha$ line seen there is not found in the \xsh spectra of UCAC2 13726137, and neither are the instances of [Fe] emission. Some metallic lines from \ion{Mg}{i} and \ion{Fe}{i} are observed in emission in Miras \citep{yao_mira_2017}, while the many \ion{Fe}{ii} emission lines that are visible in the \xsh optical spectra, are in fact consistent with semi-regular variable spectra \citep{jaschek_behavior_1995}. However, the emission line spectrum does show lines that require a hot ionising plasma, such as \ion{He}{ii} at 4686\,\AA, which cannot simply be explained by intrinsic variability of the supergiant. The corresponding lack of X-ray variability cannot be explained at the moment. 

Overall, we conclude that emission lines generated by intrinsic variability similar to what is seen in Miras and semi-regular variables could explain at least some of the lines observed in the spectra of UCAC2 13726137, considering that some of the higher ionisation lines typically observed in SySts and SyXBs are not seen here. But the \ion{He}{ii} line indicates that there is further contribution due to the presence of the compact object, and we speculate that this should correlate with the X-ray behaviour of the source. Further study and detailed modeling needs to be conducted in order to fully disentangle the origin of these emission lines, but we state here that it likely entails a contribution from intrinsic variability of the donor star, in addition to photo-ionisation from the X-ray source.

\begin{table}
\centering
  \caption{The four subclasses of semi-regular variables and how they are classified.}
  \renewcommand{\arraystretch}{1.4} 
  \renewcommand{\tabcolsep}{1.2mm}
  \begin{tabular}{ccccc}
    \hline \hline
    SRV & Spectral & Luminosity & Periodicity & Amplitude \\
    subtype & type & class & days & mag \\
    \hline
    SRa & M(e), C(e), S(e) & III & 35--1200 & <2.5 \\
    SRb & M(e), C(e), S(e) & III & 20--2300 & -- \\
    SRc & M(e), C(e), S(e) & I or III & 30--1000s & ${\sim}1$\\
    SRd & F, G, K & I or III & 30--1100 & 0.1--4 \\
    \hline
  \end{tabular}
  \label{tab:SR_class}
\end{table}

\section{Characterising the X-ray source}\label{section:x-ray_analysis}

Having discussed the donor star in the previous section, we now turn to a discussion of the properties of the X-ray source. We start in Sect.~\ref{subsec:timing} with a study of the variability behavior, followed by a discussion of the X-ray spectral properties in Sect.~\ref{subsec:spec_analysis}.

\subsection{Timing analysis}
\label{subsec:timing}
\begin{figure}
  \centering
    \includegraphics[width=\linewidth]{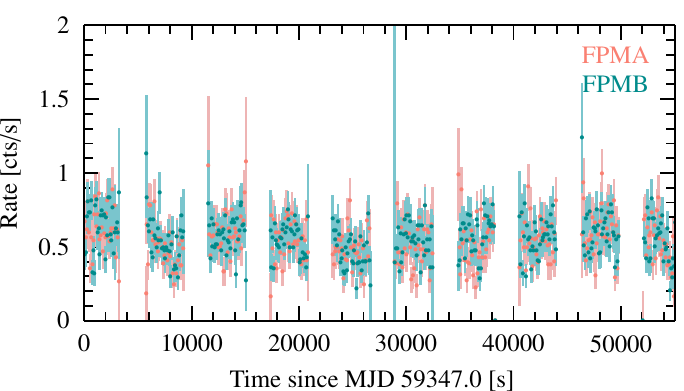}
    \caption{100s binned lightcurve of \srcname. The lightcurve shows barely any variability within uncertainties.} 
    \label{fig:lc}
\end{figure}

The underlying physics of an X-ray binary is driven by the nature of its
compact object, which in most HMXBs is typically a neutron star and not a
black hole \citep{2023A&A...671A.149F}.  
The most reliable way to confirm the presence of a neutron
star is through the detection of pulsations \citep[e.g.,][]{2004A&A...423..301T, 2009A&A...494.1073R}.

The \textit{NuSTAR} FPMA and FPMB lightcurves show the system to be in a state of relatively constant flux across the ${\sim}$55\,ks
long observation, as Fig.~\ref{fig:lc} shows. We searched for any period in the 0.1 to
6000\,s band using epoch folding \citep{1983ApJ...272..256L} and a
$Z^2$ search \citep{1983A&A...128..245B,2011ApJ...731..131G}, as well
as standard Fourier analysis and the Lomb Scargle periodogram
\citep{1982ApJ...263..835S,1986ApJ...302..757H}. Searches
were performed for the whole energy range of the detector, 3--79\,keV,
as well as for the 3--15\,keV and 15--60\,keV bands. We note that \textit{NuSTAR's} orbit results in gaps in the
lightcurves every $\sim$90\,minutes, leading to aliasing effects in
the light curve analysis. 

\begin{figure}
    \centering
    \resizebox{\hsize}{!}{\includegraphics{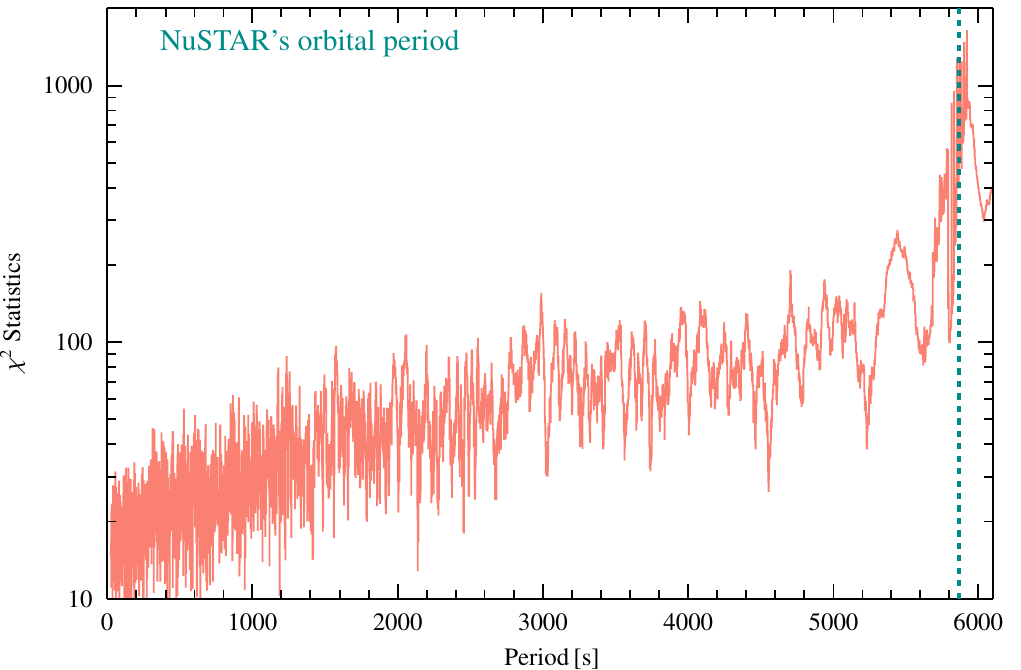}}
    \caption{Epoch folded periodogram of barycentred event file from \textit{NuSTAR}'s FPMA detector. The signal around ${\sim}$5800\,s is most likely an effect of \textit{NuSTAR}'s orbital period.} 
    \label{fig:lomb_scargle}
\end{figure}

As an example of the analysis, Fig.~\ref{fig:lomb_scargle} shows the
epoch-folded events from detector FPMA in the range 1--6000\,s, using the entire energy range of \textit{NuSTAR}. Except for a
peak at ${\sim}5800$\,s, or 96\,minutes, a signature of \textit{NuSTAR's} orbital period, no significant period is found. At
lower frequencies, low statistics of the data and the
dominance of \textit{NuSTAR}'s orbit prevent robust detection of a
signal. We also used the $Z^2$-periodogram on event data to
eliminate possible aliasing effects due to binning, to search for
periods in the range from 0.01 to 600\,s. 
Again, no significant
periodicity was found. 

In order to determine an upper limit on the amplitude of any
undetected signal that may be present, we used the analytical approach
of \citet{1994MNRAS.268..709B}. Beyond 200\,s, the data are dominated
by noise, and the method is unable to compute reliable pulsed
fractions at this point. For the range 1--200\,s, the pulsed fraction
3$\sigma$ upper limit on a sinusoidal signal is 3.56\%, in the 3--79\,keV energy range. Pulse periods in this range simulated with \textit{NuSTAR}'s
time resolution and the estimated pulsed fraction appeared above the
detection level. 

\subsection{Spectral analysis}
\label{subsec:spec_analysis}
\subsubsection{Individual eROSITA spectra}
While a detailed spectral analysis 
is not possible for the
eROSITA spectra due to the low signal to noise ratio, we use
the eROSITA data to estimate the source flux by fitting them
with an absorbed power-law, applying Cash statistics
\citep{1979ApJ...228..939C}, and the absorption model of
\citet{2000ApJ...542..914W} with abundances therein, and cross sections from
\cite{1996ApJ...465..487V}. The resulting goodness of fit varies from case to case, with reduced C-statistics in the range
0.5--0.8, due to low counts. Moreover, the photon index is negative in two of the four cases, and is around ${\sim}$0.5 otherwise. The column density is in the range $N_\mathrm{H}{\sim}1$--$2\times10^{22}\,\mathrm{cm}^{-2}$. We froze the photon index at 0.5, to determine fluxes. Over the two years of observations the system shows no decrease in flux within uncertainties in the 0.2--10.0\,keV range. This suggests no or, at most, marginal, variability, the average unabsorbed flux being ${\sim}7.75\times10^{-12}\,\mathrm{erg}\,\mathrm{cm}^{-2}\,\mathrm{s}^{-1}$. Adopting the ${\sim}$7.45\,kpc distance, \srcname displays a stable luminosity of $\sim5.14\times10^{34}\,\mathrm{erg}\,\mathrm{s}^{-1}$ in the eROSITA band, during the scans. 

\subsubsection{Describing the broadband continuum}

In order to get a better handle on the overall spectral shape than from
eROSITA alone, we next turn to the analysis of the joint
\textit{NuSTAR} and eROSITA data cumulated over the four
eRASS scans. The spectra were rebinned to at least 25 counts in each bin.

We begin the spectral analysis by modeling the 0.2--10.0\,keV
eROSITA spectrum and the 3--60\,keV \textit{NuSTAR} FPMA and
FPMB spectra together with models typically used to describe the
continuum of accreting binaries, including an absorption
component \texttt{TBabs} \citep{2000ApJ...542..914W} to
account for Galactic absorption in all models, again adopting the cross sections from
\citet{1996ApJ...465..487V} and abundances from
\citet{2000ApJ...542..914W}. The \textit{NuSTAR} energy range is chosen as such, since background dominates from 60\,keV.

First, we use a simple power law model with a folding energy, \texttt{cutoffpl} in XSPEC-like notation, 
\begin{equation}\label{eq:cut}
\centering F(E) = K \times E^{-\Gamma}\exp\left( \frac{-E}{E_{\mathrm{fold}}} \right),
\end{equation}
where $F(E)$ is the photon flux, $\Gamma$ is the photon index,
$E_{\mathrm{fold}}$ is the folding energy, and $K$ is a normalisation
constant. This model is typically used to phenomenologically describe Comptonised photons from
either the polar caps of the neutron star or originating from the
black hole's corona. We use a constant multiplier to account for differences in
the flux calibration between the two detectors of \textit{NuSTAR} and eROSITA, referring all fluxes to \textit{NuSTAR}-FPMA. We include a
Gaussian emission line to the fit to model an evident
feature in the Fe\,K$\alpha$ band around 6.4\,keV. This spectral
model, hereafter Model~1a (or M1a), results in reasonable fit statistics with
$\chi^2/\mathrm{dof}=853.43/706$ and $\chi^2_{\mathrm{red}}=1.20$, resulting in a photon index $\Gamma{\sim}0.13$, on the lower end of what is observed for accreting binaries \citep{2023arXiv230802645F}. The folding energy $E_{\mathrm{fold}} \simeq 9$\,keV is within the range of what has been observed for neutron star binaries \citep[e.g.,][]{2013A&A...551A...6M, 2020A&A...643A.128K}, while lower than usual for black holes \citep{2006csxs.book..157M}. 

We also used other continuum models that are typically applied to neutron
star spectra \citep[e.g.,][]{2013A&A...551A...6M, 2022A&A...660A..19D}, such as a Fermi-Dirac
cutoff \texttt{FDcut} \citep{1986SprT} which results in reasonable fit statistics ($\chi^2/\mathrm{dof}=  795.51/51$, $\chi^2_{\mathrm{red}}=1.12$) or the negative-positive
exponential model \texttt{NPEX} \citep{1995AAS...18710403M}, which results in better fit statistics $\chi^2/\mathrm{dof}=777.43/704$, $\chi^2_{\mathrm{red}}=1.10$), but produces unconstrained parameters, with positive and negative photon indices of $\Gamma_{1}=0.56^{+0.6}_{-0.5}$ and $\Gamma_{2}=-2.1^{+0.7}_{-0.8}$.

Our spectral modeling reveals the possible presence of additional spectral
components near 8\,keV and above 30\,keV in the residuals. Considering similar residuals, we adopt the simpler and better-known \texttt{cutoffpl} for further modeling to investigate them. Our initial fit with a cutoff power-law continuum and a neutral, homogeneous absorber shows strong residuals around the Fe K edge. We find that the prominent Fe K edge and continuum absorption are well modeled by a neutral partial covering absorber, modeled using \texttt{pcfabs}. The inclusion of the partial covering absorber (M1b) results in a better fit ($\chi^2/\mathrm{dof}=  731.65/704$, $\chi^2_{\mathrm{red}}=1.04$), and is the only fit so far where the residuals do not show a dip around 8\,keV likely corresponding to an absorption edge. The corresponding photon index $\Gamma \simeq 0.5$ is higher than the previous value, perhaps hinting that the previous fit partly accounted for the absorption edge with a low photon index. However, we find that the model does not offer a good fit to the lowest eROSITA data point. Although the energy resolution and signal-to-noise ratio of the eROSITA and \textit{NuSTAR} spectra do not facilitate probing the ionization state of the absorbing medium, we replace the neutral partial covering absorber with an ionised partial covering component \texttt{zxipcf} (M1c). This indeed results in improved fit statistics ($\chi^2/\mathrm{dof}= 698.23/703$, $\chi^2_{\mathrm{red}}=0.99$) and better models the eROSITA data at the lowest energy bin. The corresponding photon index $\Gamma \simeq 0.8$, is even higher, and the model predicts lower column density for the absorber $N_{\mathrm{H}}\lesssim10^{23}\,\mathrm{cm}^{-2}$, but a higher covering fraction. The usage of \texttt{tbpcf} to model the partial covering absorber, resulted in very similar fit parameters as with \texttt{pcfabs}.

We estimate the equivalent width of the Fe\,K$\alpha$ line for the neutral partial covering model to be $\mathrm{EW}(\mathrm{Fe}_{\mathrm{K}\alpha})\simeq691\pm45\,\mathrm{eV}$, indicating strong fluorescence. Given the necessity for a strong absorber in the line of sight to account for the absorption edge evident in the spectrum, this is to be expected under the assumption that the absorbing material is also responsible for fluorescence \citep{1985SSRv...40..317I, 2011A&A...535A...9F}. 
The ionised absorber \texttt{zxipcf} models the line with a more complex profile than we can resolve with \textit{NuSTAR}, and in particular includes an Fe absorption line at 6.7\,keV, such that the Fe\,K$\alpha$ emission line is narrower ($\mathrm{EW}(\mathrm{Fe}_{\mathrm{K}\alpha})\simeq345\pm35\,\mathrm{eV}$). The two scenarios also result in different $N_\mathrm{H}$ values for the partial absorber (see Table~\ref{tab:spectral_pars}). A stringent constraint on the column density would require disentangling the ionisation state of the absorber and cannot be carried out with the data at hand. Considering that the \textit{NuSTAR} spectra cannot resolve the complex line profile modeled by \texttt{zxipcf} and that the errant eROSITA bin for Model~1b spans a wide energy range, we adopt the neutral partial covering absorber for further spectral fitting, with caveats as stated above.

We modify the continuum model (M1b) with the inclusion of \texttt{cyclabs}, a
component to model cyclotron absorption with a pseudo-Lorentzian profile, in order to probe the presence of
cyclotron resonant scattering features (CRSFs) in the spectra
\citep[see, e.g.,][]{1978ApJ...219L.105T,staubert2019}. The fit statistics and residuals are not significantly improved by the inclusion of an absorption component at around 34\,keV,
indicating that a cyclotron component likely cannot be adequately resolved here
($\chi^2/\mathrm{dof}=717.22/701$). The fit was also conducted using \texttt{gabs}, a cyclotron absorption component with a Gaussian optical depth parameter, with similar results. 

We find a discrepancy in the cross-calibration between eROSITA and \textit{NuSTAR} of up to 40\%. While part of this can be attributed to known calibration issues of the eROSITA hard band with instruments like \textit{XMM-Newton}, which has then been used to get an indirect relation for \textit{NuSTAR} by \citet{2024A&A...688A.107M}, this only amounts to about 3--5\%. Since the eROSITA and \textit{NuSTAR} observations are not contemporaneous, the difference in flux could also be source-intrinsic. Nevertheless, a discrepancy by a factor $\lesssim2$ within uncertainties, is well within typical variability observed for wind accretors \citep{kretschmar_advances_2019}. 

\begin{figure}
  \centering
    \resizebox{\hsize}{!}{\includegraphics{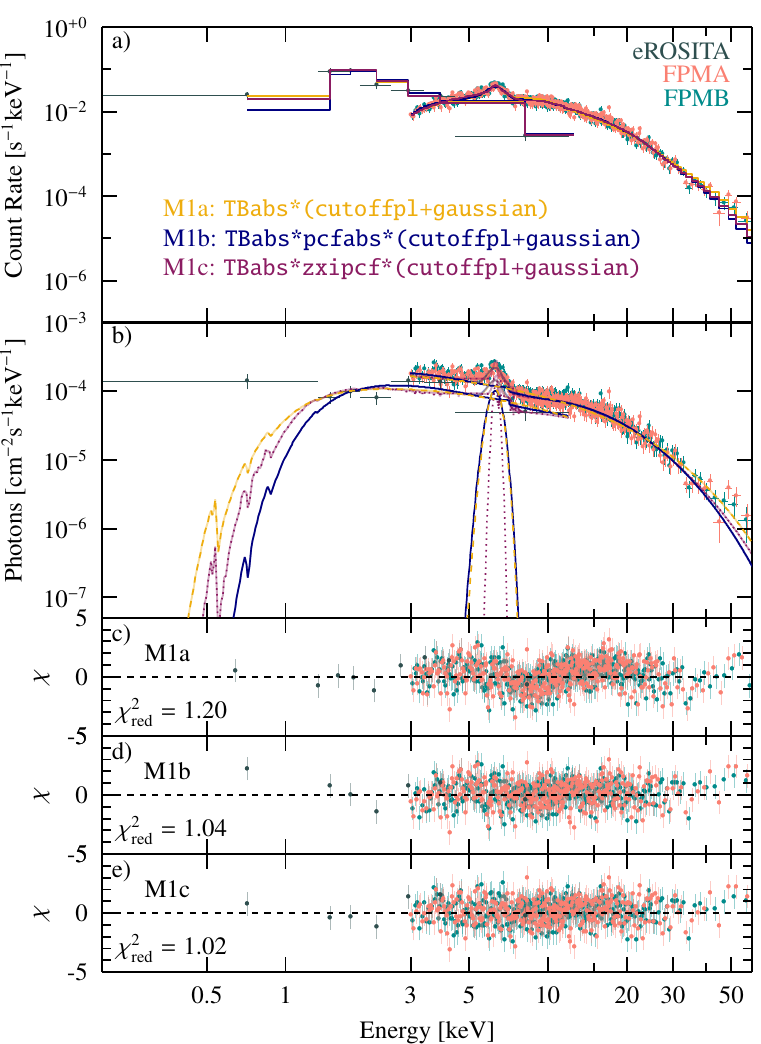}}
    \caption{\textit{NuSTAR} and eROSITA spectra of \srcname from FPMA (teal) and FPMB
      (peach) fitted with three different models. The folded and
      unfolded models with individual model components are shown in panels
      \textbf{a} and \textbf{b} respectively. The $\chi^2$-residuals
      are shown for \textbf{c} Model\,1a (M1a), the high-energy cutoff, \textbf{d} M1b, the high-energy cutoff including a partial absorber, and
      \textbf{e} M1c, the high-energy cutoff including an ionised absorber.} 
    \label{fig:spec_models}
\end{figure}

\subsubsection{``Best" spectral fits}
Our spectral fits M1a--c assume the same underlying continuum but differ in the modeling of the source-intrinsic absorption with regard to the geometrical/temporal variability and possible degree of ionization. The question of the absorption model is primarily raised by strong residuals around the Fe\,K edge left in our simplest, fully covering, neutral absorption model M1a. Diagnostics about the absorbing medium are, however, hampered because of \textit{NuSTAR}'s energy resolution and the limited soft coverage. As shown in Fig.~\ref{fig:spec_models}, the models diverge mostly below 3\,keV. Best-fit parameters are given in Table~\ref{tab:spectral_pars}, with 90\% confidence limits. 

We note that the baseline $N_\mathrm{H} (\texttt{tbabs}$) in all our evaluated models agrees within uncertainties, along with estimated intrinsic fluxes which were determined using the convolution model \texttt{cflux} after fixing the parameters to their best-fit values for each model. The unabsorbed fluxes were obtained by setting the absorption contribution to zero. The partial covering models M1b and M1c contain a higher column density component that gives rise to a deeper Fe\,K edge, therefore resolving the modeling issues between 6--8\,keV. Statistically, our analysis favors model M1c but we note that the choice of the absorption model has only a marginal effect on the derived source luminosity.

\begin{table*}
\renewcommand{\arraystretch}{1.32} 
\renewcommand{\tabcolsep}{2mm}
\caption{Spectral parameters for different models used to fit the eROSITA and \textit{NuSTAR} spectra, with $2\sigma$ confidence intervals.} 
\centering
\begin{tabular}{p{3cm}p{4cm}p{3cm}p{3cm}p{3cm}}                
 \hline                                  
 \hline            
 \multicolumn{5}{c}{Model~1: \texttt{TBabs*\textcolor{gray}{pcfabs/zxipcf}*(cutoffpl+gaussian)}}\\
 \multicolumn{2}{c}{Component} & M1a & M1b & M1c\\
 \hline                                 
\texttt{constant} & $C_\mathrm{FPMB}$ & $1.064\pm0.022$ & $1.065^{+0.022}_{-0.021}$ & $1.064^{+0.022}_{-0.021}$\\
& $C_\mathrm{eROSITA}$ & $0.62^{+0.14}_{-0.12}$ &  $0.65^{+0.16}_{-0.14}$ & $0.60^{+0.14}_{-0.13}$\\
\texttt{tbabs} & $N_\mathrm{H}$ [$10^{22}\,\mathrm{cm}^{-2}$] & $0.8^{+0.5}_{-0.4}$ & $1.8^{+1.3}_{-0.9}$ & $1.4^{+0.8}_{-0.5}$\\
\texttt{pcfabs}/\texttt{zxipcf} & $N_\mathrm{H}$ [$10^{22}\,\mathrm{cm}^{-2}$] & -- & $\left(2.4^{+0.6}_{-0.5}\right)\times10^{2}$ & $75\pm11$ \\
& $f_{\mathrm{pc}}$ & -- & $0.54\pm0.06$ & $0.72^{+0.07}_{-0.09}$\\
& $\log\xi$ & -- & -- & $2.27^{+0.19}_{-0.14}$ \\
\texttt{gaussian} & $K_\mathrm{gaussian}$ [$\mathrm{photons}\,\mathrm{cm}^{-2}\,\mathrm{s}^{-1}$]& $\left(9.3^{+1.0}_{-0.9}\right)\times10^{-5}$ &  $\left(1.95^{+0.33}_{-0.28}\right)\times10^{-4}$ & $\left(9.2^{+1.9}_{-1.6}\right)\times10^{-5}$\\
& $E_\mathrm{Fe K\alpha}$ [keV] & $6.29^{+0.04}_{-0.05}$ & $6.27\pm0.05$ & $6.34\pm0.04$\\
& $\sigma_\mathrm{Fe K\alpha}$ [keV] & $0.35^{+0.06}_{-0.05}$ & $0.37^{+0.07}_{-0.06}$ & $0.16\pm0.06$\\
\texttt{cutoffpl} & $\Gamma$ & $0.13^{+0.08}_{-0.07}$ & $0.40^{+0.19}_{-0.16}$ & $0.84^{+0.31}_{-0.28}$\\
& $K_\mathrm{cutoffpl}$ [$\mathrm{photons}\,\mathrm{cm}^{-2}\,\mathrm{s}^{-1}$] & $\left(2.70^{+0.33}_{-0.29}\right)\times10^{-4}$ & $\left(1.02^{+0.45}_{-0.26}\right)\times10^{-3}$ & $\left(2.2^{+2.1}_{-1.0}\right)\times10^{-3}$ \\
& $E_\mathrm{fold}$ [keV] & $10.8^{+0.7}_{-0.6}$ & $8.9^{+0.7}_{-0.6}$ & $11.7^{+2.3}_{-1.7}$\\
\texttt{eqwidth} & $\mathrm{EW}(\mathrm{Fe}_{\mathrm{K}\alpha})$ [$\mathrm{eV}$] & $780\pm60$ & $691\pm45$ & $345\pm35$ \\
\hline
$\chi^2/\mathrm{dof}$ & & 853.43/706 & 819.14/784 & 779.32/776  \\
\hline
\texttt{cflux} & log $f_{3-55, \mathrm{abs}}$\,[$\mathrm{erg}\,\mathrm{s}^{-1}\,\mathrm{cm}^{-2}$] & $-10.493\pm0.005$ & $-10.504\pm0.005$ & $-10.497\pm0.005$\\
 & log $f_{3-55, \mathrm{unabs}}$\,[$\mathrm{erg}\,\mathrm{s}^{-1}\,\mathrm{cm}^{-2}$] & $-10.483\pm0.005$ & $-10.328\pm0.005$ & $-10.338\pm0.005$\\ 
& $L_{3-55,\mathrm{unabs}}$\,[$\mathrm{erg}\,\mathrm{s}^{-1}$] & $2.18\times 10^{35}$& $3.11\times 10^{35}$& $3.04\times 10^{35}$\\
\hline
\end{tabular}
\label{tab:spectral_pars}
\end{table*}

\section{Discussion}
\label{section:discussion}
The results of the previous sections can be summarized as follows:
\begin{itemize}
    \item \srcname was detected in each eROSITA scan showing persistently bright flux.
    \item The spectroscopic and photometric data of the optical counterpart reveal an M2--3 supergiant donor, with several transient emission lines. This indicates that the X-ray binary in question is an HMXB, albeit an unusual one.  
    \item The persistent nature of \srcname and its late-type supergiant companion place it in a sparsely populated class of only two in the Galaxy. 
    \item The hard photon index, cutoff energy, and luminosity all favor a neutron star as the accreting object. 
   \item We do not detect pulsations in the \textit{NuSTAR} lightcurves, but suggest that pulsations longer than 200\,s would not be detected due to the low sensitivity of the observation to pulsations in that period range. Such larger spin periods are commonly seen in X-ray binaries with late-type companions, or obscured HMXBs. 
    \item The detection of a strong Fe\,K$\alpha$ line with $\mathrm{EW}(\mathrm{Fe}_{\mathrm{K\alpha}})\sim700\,\mathrm{eV}$ is indicative of obscuration of the X-ray source by the wind of the supergiant companion or other circumstellar material.
    \item The presence of circumstellar material in the system is also consistent with a red supergiant companion that shows intrinsic variability. 
\end{itemize}
In the following, we discuss 
each of the above points in more detail.

\subsection{Persistent X-ray luminosity}
\label{subsec:gaia_lum}
We use the estimated geometric distance of $d =7.45^{+0.75}_{-0.71}$\,kpc, as obtained from \citet{2021AJ....161..147B} for our luminosity estimates\footnote{\citetalias{2024MNRAS.528L..38D} adopt a value of ${\sim}12$\,kpc using the Gaia parallax, making our luminosity estimates not directly comparable.}. For completeness, the photogeometric distance is $d =6.58^{+0.38}_{-0.44}$\,kpc \citep{2021AJ....161..147B}.

The unabsorbed 0.2--55\,keV luminosity of \srcname, estimated using the
adopted distance of 7.45\,kpc and the model assumptions discussed in Sect.~\ref{subsec:spec_analysis}, is
${\sim}2.1$--$3.1\times 10^{35}\,\mathrm{erg}\,\mathrm{s}^{-1}$. As shown in Fig.~\ref{fig:lum-states}, the flux state is consistent between the eROSITA scans. A similar luminosity is reported by \citetalias{2024MNRAS.528L..38D}, using the \textit{Swift/XRT} data, consistent within uncertainties despite the difference in the adopted distances. Although a drop in flux between these observations would not have been noticed, the fact that each detection of \srcname finds it at the same flux lends credence to its higher duty cycle at this flux state.

Further, the \textit{Swift}/BAT lightcurve of \srcname displays consistently low flux $\lesssim$2\,mCrab for its entire duration of 13\,years, corresponding to an average 14--195\,keV flux of ${\sim}1.2\times10^{-11}\,\mathrm{erg}\,\mathrm{cm}^{-2}\mathrm{s}^{-1}$. The lack of \textit{Swift}/BAT detections of any outbursts from \srcname over 13\,years supports its persistent nature at ${\sim}10^{35}\,\mathrm{erg}\,\mathrm{s}^{-1}$. Moreover, \srcname displays consistent flux levels between \textit{Swift/BAT} and the eROSITA snapshots and \textit{NuSTAR} observation nearly 4\,years later. The non-detections with \textit{ROSAT} and \textit{INTEGRAL} are also consistent with this low flux (see Table~\ref{tab:survey_data}), placing this source closer to the realm of persistent HMXBs than transients.

\subsection{The optical picture: spectral type and emission lines}

As detailed in Sect.~\ref{section:optical}, we determined the spectral type and luminosity class of \srcname to be an M2--3 type supergiant. We used TiO indices defined by \citet{kenyon_cool_1987} to compute the spectral type, finding that a K type scenario is inconsistent with the strength of TiO bands at 6180\,\AA\xspace and 7100\,\AA\xspace, among others, in the optical and NIR \xsh spectra. We then used the absorption features of \ion{Ti}{i}, \ion{Fe}{i}, CN, the first CO overtone  $\mathrm{CO}(2,0)$, and other diagnostics also used by \citetalias{2024MNRAS.528L..38D} to confirm that the optical counterpart to \srcname is indeed a late-type supergiant, as suggested by \citetalias{2024MNRAS.528L..38D}. This makes \srcname only the second known Galactic X-ray binary with a red supergiant companion, in addition to 4U~1954$+$31 \citep{hinkle_m_2020}.  

We then investigated the likely cause of the many emission lines present in the \xsh spectra of UCAC2~13726137, finding them similar to those present in variable stars like Miras \citep{wood_mira_1990,castelaz_phase-dependent_2000} or the less well defined class of SRVs \citep{jaschek_behavior_1995}. Alternate scenarios where the emission lines are solely due to accretion were found insufficient to explain the whole picture since the \xsh spectra do not show the many high ionisation state emission lines that are typically observed in the spectra of SyXBs like GX~1$+$4 \citep{davidsen_optical_1977} or IGR~J17329$-$2731 \citep{bozzo_igr_2018}. Instead, most of the lines we observe in emission have been observed in variable stars (see Sect.~\ref{section:optical} for details). 

Nevertheless, the presence of the \ion{He}{ii} $\lambda4686$ line in emission, indicates a hard ionizing source in the system. \ion{He}{ii} has been associated with an accretion disk around the compact companion, in a handful of cases \citep{2023arXiv230802645F}. However, Roche Lobe Overflow (RLOF) is thought to be the predominant accretion mode in these systems \citep[][and references therein]{kretschmar_advances_2019}, which we do not consider likely for \srcname. Transient accretion disks which have been observed in some HMXBs \citep{sidoli_integral_2018} could still explain the presence and intrinsic variability of the \ion{He}{ii} emission lines, although this also comes with periods of enhanced luminosity. \ion{He}{ii} emission has also been associated with nebular emission in a handful of symbiotic stars \citep{merc_new_2019}. In comparison to \ion{He}{ii}, there is much more variability in the H$\alpha$ emission line (see Fig.~\ref{fig:halpha_line}). Since this is difficult to reconcile with the low variability of the X-ray source, we suggest that it could arise from the late-type companion, as has been observed in variable M giants. The caveat is that there is much less precedence for supergiant systems with similar optical spectra, although there are certainly some examples \citep{levesque_physical_2009}. A scenario where the compact object is in a low-eccentricity orbit around the massive star in order to allow for persistent accretion where the shape and intensity of the emission lines in the optical spectra are primarily due to geometric effects such as inclination and viewing angle is not out of the question, but necessitates a monitoring campaign in both optical and X-ray regimes to confirm. 

\subsection{Evolutionary track and likely accretion mode}
Although there is precedence for a binary system with a compact object accreting from a red supergiant \citep{hinkle_m_2020}, we attempt to trace such a system's journey by considering typical binary evolution scenarios. There are three scenarios that can result in a system hosting an RSG and a compact object \citep[e.g.,][their Fig.~1]{han_binary_2020}. In each case, the binary system starts off consisting of two massive stars and the preceding stage to the RSG phase is a wind-accreting HMXB phase. One of the scenarios involves a common envelope phase which then results in a binary system with a Helium star and a main sequence star. But in each case, there is a stage of mass transfer where a He star accretes from a main sequence companion, eventually undergoing a supernova explosion and resulting in a HMXB with wind accretion. Depending on its initial mass, after leaving the main sequence, the massive companion star becomes a red supergiant resulting in this particular stage of binary evolution. This phase is comparably short since the RSG phase is marked by a significant amount of mass loss, resulting in a second supernova, eventually giving rise to a double compact binary. Although this is not explained in further detail by \citet{han_binary_2020}, RSG+CO phase is assumed to have mass transfer via RLOF, which, as mentioned above, we do not observe any indication of. We therefore assume that the primary mode of mass transfer is wind accretion. 

\citetalias{2024MNRAS.528L..38D} suggested that the observed X-ray luminosity of \srcname implies unrealistically large binary separations of $a\sim25000\,R_{\odot}$. We use the same accretion efficiency coefficient, $\eta=0.2$, and using $L_{X}=\eta\dot{\mathrm{m}}c^2$ \citep{longair_high_2011}, and obtain a mass accretion rate onto the compact object of $\dot{M}_\mathrm{NS}=1.72\times10^{15}\,\mathrm{g}\,\mathrm{s}^{-1}$, adopting an X-ray luminosity of $3\times10^{35}\,\mathrm{erg}\,\mathrm{s}^{-1}$. Under the assumption of Bondi-Hoyle-Lyttleton accretion \citep{1944MNRAS.104..273B,edgar_review_2004} and a circular orbit around the donor star, 
\begin{equation}
    \dot{M}_{\mathrm{NS}} = \frac{(GM_{\mathrm{NS}})^2}{a^2v_{\mathrm{rel}}^{3}v_{\mathrm{wind}}}\dot{M}_{\mathrm{wind}},
\end{equation}
where $G$ is the Gravitational constant, $M_{\mathrm{NS}}$ is the mass of the neutron star, which we assume to be $1.4\,M_{\odot}$, it is possible to obtain reasonable binary orbital separation $a\lesssim2000\,R_{\odot}$, for a mass loss rate of $\dot{M}_{\mathrm{wind}}=10^{-8}M_{\odot}\,\mathrm{yr}^{-1}$ \citep{beasor_new_2020}, adopting a slightly higher mass of the red supergiant \citep[see][for a wide range of possible radius and mass values]{redsupergiants}. The relative velocity $v_\mathrm{rel}$ requires taking into account both the orbital velocity of the neutron star and the stellar wind velocity $v_{\mathrm{wind}}$, for which we assume a typical value of about ${\sim}20\,\mathrm{km}\,\mathrm{s}^{-1}$ \citep{goldman_wind_2017}, as did \citetalias{2024MNRAS.528L..38D}. 

\subsection{The compact object}
\label{subsec:compact-object}

X-ray emission arising from Comptonisation occurring near the compact object's surface or corona, phenomenologically modeled using \texttt{cutoffpl}, describes the data well, and would be in line with both neutron star and black hole binary spectra \citep{1994ApJS...92..511V, 2006ChJAS...6a.183D}. However, the photon index $\Gamma\simeq 0.5$ is significantly harder than typical values for black hole binaries \citep{2006csxs.book..157M, 2019ApJ...879...93X,2020MNRAS.493.5389F}, for which $\Gamma{\sim}1.5$--2.0 range. The low folding energy, $E_\mathrm{fold}\simeq8\,\mathrm{keV}$, is also rather unusual for black hole binaries where the exponential cutoff is expected between 30--100\,keV \citep{2006csxs.book..157M}, while neutron star binaries do show similar values \citep[see Fig.~1.9 of][]{Falkner2018}. 

The observed X-ray luminosity of the order of a few $10^{35}\,\mathrm{erg}\,\mathrm{s}^{-1}$ is atypical for black hole binaries \citep{2023ApJ...944..165B}, which appear at much lower luminosities ${\lesssim}10^{32}\,\mathrm{erg}\,\mathrm{s}^{-1}$ when in quiescence \citep[e.g.,][]{2022A&A...661A..20C}, if even detected. Further, evolutionary predictions for binary systems hosting red supergiants with black hole companions imply that this stage of binary evolution is extremely short-lived and is very unlikely to be observed \citep{klencki_it_2021}.  

A white dwarf accretor should also be considered, since the optical spectra show striking resemblance to those of SySts (white dwarfs accreting from late-type giants). The vast majority of SySts have soft X-ray spectra \citep{luna_symbiotic_2013}, like StH$\alpha$\,32 \citep{orio_two_2007}, but there have also been detections of sources with harder X-ray emission, even exceeding 20\,keV like RT\,Cru \citep{2020mbhe.confE..45C,tueller_swift_2005}. The corresponding X-ray spectra are also highly absorbed and it is speculated that the hard X-ray emission arises at the truncation radius of the accretion disk \citep{luna_symbiotic_2013}. However, the relatively higher luminosity of \srcname, also rules out a white dwarf as the compact object. The maximum X-ray luminosities observed for white dwarf symbiotic systems are only up to a few times ${\sim}10^{33}\,\mathrm{erg}\,\mathrm{s}^{-1}$ \citep{luna_symbiotic_2013}, while \srcname is brighter by a factor of ${\sim}100$. 
 
The handful of other X-ray binaries where the donor is a red supergiant -- 4U~1954$+$31 \citep{hinkle_m_2020} and the extragalactic NGC\,300\,ULX-1 \citep{heida_discovery_2019} -- both contain neutron stars for which pulsations have been detected. This is too limited a sample to serve as a convincing argument, but we conclude on \srcname most likely containing an accreting neutron star compact object by exclusion, following the preceding discussion of alternative scenarios. Further, based on evolutionary tracks and characteristic ages of red supergiants, the accreting neutron star should be highly magnetised with magnetic fields of the order of a few $10^{12}\,\mathrm{G}$ \citep{TaurisvandenHeuvel+2023+376+432, han_binary_2020}.

\subsection{Comparison to 4U~1954$+$31 and other HMXBs}
\label{subsec:comparison}
As an X-ray binary with a red supergiant donor suspected to host a neutron star, \srcname merits comparison with the only other Galactic system known to fit this configuration, 4U~1954$+$31. 4U~1954$+$31 has long been studied as a SyXB \citep{enoto_spectral_2014,2011ApJ...742L..11M} since it was thought to contain an M~giant donor, until revised optical identification by \citet{hinkle_m_2020} revealed an M~supergiant donor instead. It has been observed at similar X-ray luminosities to \srcname, but also shows flaring behaviour \citep{enoto_spectral_2014,bozzo_accretion_2022}, reaching upto luminosities of $10^{36}\,\mathrm{erg}\,\mathrm{s}^{-1}$. The broadband X-ray spectrum has been modeled with a variety of spectral models over the years \citep{2011ApJ...742L..11M,enoto_spectral_2014,bozzo_accretion_2022}, but is typically found to have a hard spectrum with an Fe K$\alpha$ emission line, and often requires a partial covering absorber.  

Another candidate is CXO~174528.79$-$290942.8, which was discovered as a result of infrared identification of X-ray binaries in the Galactic Center, and was proposed to contain a red supergiant donor by \citet{gottlieb_rapidly_2020}. Although the NuSTAR observation of the source only yielded ${\sim}$400 counts, spectral parameters reported indicate a very hard spectrum with a photon index, $\Gamma\sim0.7$ \citep{hong_nustar_2016}, similar to the X-ray spectrum of \srcname. 

Since X-ray binaries with red supergiant companions constitute a very small fraction of HMXBs, they have so far either been considered part of SyXBs \citep{de_massive_2022,2024MNRAS.528L..38D} or SgXBs \citep{hinkle_m_2020,bozzo_accretion_2022}. However, based on binary evolution scenarios \citep[as summarised by][]{han_binary_2020}, it is evident that X-ray binaries hosting a compact object accreting from an RSG, and those accreting from M~giants or Asymptotic Giant Branch (AGB) stars, follow two completely separate evolutionary tracks. The M~supergiant donors of \srcname and 4U~1954$+$31 imply vastly different mass loss rates and orbital parameters from those with M giant donors, and similarly for O/B-type supergiant donors, such that classifying them under either subclass might be misleading. There are, however, common features between these systems in the context of X-ray observations, especially since persistent wind accretion and high obscuration due to clumpy wind are well-studied phenomena in SgXBs, and since the expected magnetic field estimates for the accreting neutron star are similar for each subclass. A caveat is that there is only one robust detection of a magnetic field estimate among known SyXBs, that of IGR~J17329$-$2731, which was reported by \citet{bozzo_igr_2018} to be ${\sim}2.4\times10^{12}$\,G, but they are generally suspected to host highly magnetised neutron stars owing to their long pulse periods \citep{lu_population_2012}. Nevertheless, since the sample of sources with a similar configuration to \srcname is rather limited, and since the underlying origin of the X-ray emission is supposed to be from the vicinity of an accreting neutron star, we attempt to contrast the observations against a wider variety of source types where this assumption holds. 

The \textit{NuSTAR} spectrum of \srcname deviates from the typical spectra of Sy/SgXBs, which exhibit sharp cutoffs \citep[e.g.,][]{2005SSRv..120..143B, 2019ApJ...873...62H}, as opposed to the high energy flattening and rising of the spectra seen in low luminosity BeXRBs \citep{2021A&A...651A..12S}. Both examples are shown in Fig.~\ref{fig:low_lum}, where we compare the \textit{NuSTAR} spectrum of \srcname with those of a handful of different HMXBs, X~Persei \citep{2012A&A...540L...1D}, the aforementioned 4U~1954$+$31, two SgXBs 4U~1700$-$37 \citep{bala_possible_2020} and 4U~1538+22 \citep{2019ApJ...873...62H}, the SyXB IGR~J17329$-$2731 \citep{bozzo_igr_2018}, the persistently wind accreting system GX~301$-$2 with a hypergiant companion \citep{2014MNRAS.441.2539I, zalot2024} and the SFXT IGR~J16418$-$4532 \citep[][where the same \textit{NuSTAR} observation is studied for timing results]{islam_investigating_2023}. The spectral shape of \srcname notably does not match that of 4U~1954$+$31, with the absorption severely modifying the soft spectrum and the latter's cutoff at high energies. On the other hand, 4U~1954$+$31 has a similar spectrum to IGR~J16418$-$4532. The high obscuration is present in both SgXBs and SyXBs. Overall, the spectra in Fig.~\ref{fig:low_lum} serve to convey the inhomogeneous nature of the X-ray spectra of HMXB subclasses, while commonality can be found between different types, perhaps hinting at the need for something in addition to subclass as a discerning factor, while affirming the relative uniqueness of \srcname as an HMXB. 

\subsection{Low luminosity persistent XRB?}
\label{subsec:lowlum}
\citetalias{2024MNRAS.528L..38D} proposed that \srcname is in the propeller regime, based on the donor's estimated mass loss rate that they deem incompatible with the observed X-ray luminosity. Following \citetalias{2024MNRAS.528L..38D}'s results, \citet{afonina_probing_2024} explored the possibility of \srcname being in the propeller state by applying spin-down models and concluded that such a propeller stage would be too short an evolutionary phase to provide a high enough probability to observe such a system. 

Spectral signatures of the propeller regime have not yet been established in literature. The sources where centrifugal inhibition of accretion has been suggested were reported to show a spectral transition at the low luminosity state, with the spectrum tapering off around 5--6\,keV \citep[][although this is difficult to disentangle from simply low counting statistics]{tsygankov_x-ray_2017}, which is inconsistent with the \textit{NuSTAR} spectrum of \srcname, where we observe hard X-ray emission up to 55\,keV. On the other hand, a scenario where accretion is allowed to a certain extent, as is in the case of magnetic inhibition \citep[first described as a subsonic propeller by][]{davies_spindown_1979,davies_accretion_1980}, has been observed in many HMXBs, including the SFXTs \citep{shakura_wind_2015}. The case where some matter can still penetrate onto the surface of the magnetic field is certainly compatible with the observed hard X-ray emission and luminosity of \srcname, and is the only model suggested by \citet{afonina_probing_2024} to be considered realistic to \srcname. 

The estimated luminosity of ${\sim}2$--$3\times10^{35}\,\mathrm{erg}\,\mathrm{s}^{-1}$ (see Sect.~\ref{subsec:gaia_lum}) is consistent with that of several HMXBs -- wind-accreting SgXBs such as GX~301$-$2, 4U~1538$-$52, 4U~1909$+$07, and IGR~J19140$+$0951 \citep[see][and references therein]{sidoli_integral_2018} to name a few, as well as some BeXRBs such as GX~304$-$1, GRO~J1008$-$57, and A0535$+$262 \citep{2018A&A...620L..13R, 2021ApJ...912...17L, 2019MNRAS.487L..30T}, which have been detected at comparable luminosities outside of outburst. SFXTs, which are characterised by very low quiescent luminosities of $10^{32}$--$10^{33}\,\mathrm{erg}\,\mathrm{s}^{-1}$ and high dynamic ranges of $10^{2}$--$10^{5}$ \citep{sidoli_supergiant_2013}, are also detected at intermediate luminosities of $10^{35}\,\mathrm{erg}\,\mathrm{s}^{-1}$ \citep{romano_monitoring_2009,drave_supergiant_2013}. Finally, the red-SgXB, 4U~1954$+$31 has been observed at luminosities of $10^{33}$--$10^{35}\,\mathrm{erg}\,\mathrm{s}^{-1}$ \citep{enoto_spectral_2014, bozzo_accretion_2022}. 

The above examples serve to emphasise that stable accretion onto a highly magnetised neutron star as found in HMXBs, at low and intermediate luminosities has been observed across all subclasses. However, there are only a few cases where the luminosity remains stable without flaring or outbursting activity. While this might be an observational effect, it is still surprising in comparison to sources which show dynamic ranges of at least a factor ${\sim}$10 even with similarly low cadence of observations \citep{sidoli_integral_2018}. This picture of so-called `low luminosity' accretion is a matter of current study within each subclass. \citet{2017A&A...608A..17T} proposed accretion from a cold non-ionised disk to explain ongoing quiescent (i.e., outside of outburst) accretion in BeXRBs. For SFXTs, the physical scenario proposed is the settling accretion regime, such that the slowly rotating neutron star accretes quasi-spherically \citep{shakura_theory_2012,shakura_2018}, and is inhibited by a quasistatic shell above the magnetosphere at low mass accretion rates, due to inefficient cooling of the infalling matter, leading to a reduction in overall X-ray luminosity \citep{bozzo_supergiant_2015,sidoli_capturing_2023}. A similar consideration is applied to SyXBs, with accommodations for the lower wind velocities of the M~giant donors \citep{yungelson_wind-accreting_2019}. 

\subsection{The Fe\,K$\alpha$ line and obscuration}
The spectra necessitate a partial covering absorber for an acceptable spectral fit, with an $N_{\mathrm{H}}$ between ${\sim}75$--$240\times 10^{22}\,\mathrm{cm}^{-2}$. Physical scenarios producing such an effective partial covering absorption are plausible and include for example short-term variability of the absorption column that is averaged in a longer observation, or radiation scattered into the line of sight via different paths through the absorbing medium. However, quantitative interpretation of these parameters, including the ionization parameter, requires several assumptions about the composition, geometrical dimensions and density of the absorber, that cannot be validated reliably given our limited knowledge about the source and quality of the observational X-ray data. 

 The obscuration may be interpreted as due to the compact object being embedded in a stream of stellar wind from the companion star. This is a scenario often invoked for classical SgXBs, some examples of which are IGR~J16320$-$4751 \citep{2003A&A...411L.427W}, IGR~J16393$-$4643 \citep{2003A&A...407L..41R}, IGR~J17252$-$3616 \citep{2006A&A...448..261Z}, all classified as highly obscured HMXBs after detection in the \textit{INTEGRAL} survey. Consequently, at comparable luminosities, several SgXBs have strong Fe\,K$\alpha$ emission lines in their X-ray spectra, with EW(Fe\,K$\alpha$) reaching well beyond 1\,keV, provided the corresponding column density is sufficiently high \citep[][and references therein]{2018A&A...610A..50P}. The observed absorption for \srcname is still about an order of magnitude higher, and is comparable to the highly obscured IGR~J16318$-$4848 \citep{2004ApJ...616..469F, 2020A&A...641A..65B}, which has an sgB[e] companion. Although the stellar wind properties differ vastly between early and late-type supergiants, there is an evolutionary link between sgB[e] stars and red supergiants, as the former are considered to either be in a pre-RSG phase or a post-RSG phase \citep{wit_dusty_2014}, with circumstellar dust being a common property between the two stages of massive star evolution \citep{stencel_far_1987}. The presence of dusty clumpy winds might explain the high obscuration in the system, but more information on the absorber can only be obtained from a longer, high-resolution soft X-ray observation, which is outside the scope of this study. 

\begin{figure}
    \centering
    \resizebox{\hsize}{!}{\includegraphics{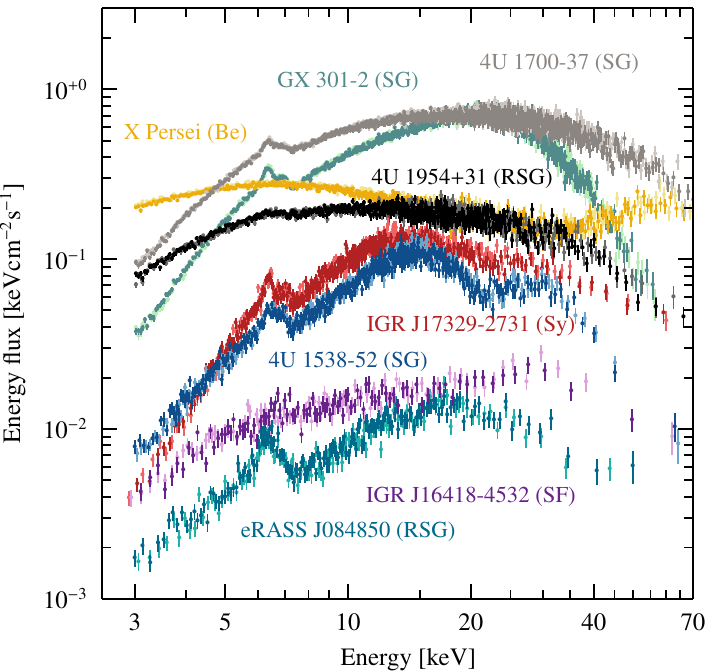}}
    \caption{X-ray spectrum eRASSU \srcname (teal) shown along with \textit{NuSTAR} spectra of a few other HMXBs for comparison: X~Persei, a persistent BeXRB displaying low luminosity accretion (yellow) \citep{2012A&A...540L...1D}, two SgXBs 4U~1700$-$37 (grey) \citep{bala_possible_2020} and 4U~1538$+$22 (blue) \citep{2019ApJ...873...62H}, a SyXB IGR~J17329$-$2731 (red) \citep{bozzo_igr_2018}, the persistent, albeit brighter wind-accretor GX~301$-$2 (light green) \citep{2014MNRAS.441.2539I, zalot2024} and the SFXT IGR\,J16418$-$4532 (purple) \citep{islam_investigating_2023}.
    FPMA and FPMB spectra have been illustrated using different shades of the same colour. The strong iron line in \srcname is comparable to that of the obscured systems and one of the SyXBs. These spectra each correspond to luminosities of 10$^{35}\,\mathrm{erg}\,\mathrm{s}$ with the exception of GX~301$-$2 which is brighter by a factor ${\sim}$10.} 
    \label{fig:low_lum}
\end{figure}

\subsection{Non-detection of pulsations}
Although the high likelihood of \srcname to be a neutron star has been established, the non-detection of pulsations must be addressed. The absence of detectable pulsations is not markedly unusual, since not all known neutron star HMXBs show pulsations \citep[see catalogs of][]{2023A&A...677A.134N, 2023A&A...671A.149F}. This is in particular the case for sources where heavy obscuration by surrounding material is detected, such as 4U~2206$+$54 \citep{2009A&A...494.1073R}, whose long pulse period was detected several decades after its initial discovery. Wind-fed obscured SgXBs IGR~J16320$-$4751 and OAO~1657$-$415 are other examples of sources where pulsations are smeared out by high levels of absorption \citep{2023NewA...9801942V,pradhan2014}. 

The other Galactic red-SgXB, 4U~1954$+$31 has a notoriously long spin period of 5.4\,h \citep{corbet_comparison_2008, enoto_spectral_2014}. The ULX with a red supergiant donor, NGC~300, however was detected to have a rather short spin period of $\sim32$\,s \citep{heida_discovery_2019}. Nevertheless, other HMXBs that spend most of their duty cycle at a comparable luminosity state like those discussed in Sect.~\ref{subsec:lowlum}, have spin periods between 100--10000\,s. In fact, those presented in Fig.~\ref{fig:low_lum} have spin periods ranging from 526\,s to 5.4\,h \citep[in ascending order of spin period,][]{davison_binary_1977, white_periodic_1976, white_x-ray_1976, bozzo_igr_2018, corbet_comparison_2008}. 
We did not detect a significant signal in this range (see Fig.~\ref{fig:lomb_scargle}). Therefore, the non-detection of a long rotation period that may be present as a weak signal in the data cannot be ruled out.

\citetalias{2024MNRAS.528L..38D} suggest that \srcname is in the propeller regime and should correspondingly spin with a very short pulsation period. As shown in Sect.~\ref{subsec:timing}, however, we do not detect any significant pulsations in this range.

\section{Conclusions}
\label{section:conclusion}
In conclusion, \fullname is among the new X-ray binaries to be studied in detail as a result of X-ray surveys of the entire sky, first by \citetalias{2024MNRAS.528L..38D} based on the \textit{Swift}/BAT survey, followed by this work, where the effort was motivated by follow-up of bright sources detected by eROSITA. Along with the discovery of SRGA~J124404.1, the discovery of \srcname possibly hints at the advent of a class of low luminosity persistent X-ray binaries whose detection is facilitated by eROSITA. We suggest that \srcname is a neutron star binary, given its
spectral parameters and the spectral type of the companion star, which we confirm to be a late-type supergiant, in accordance with De+24, although with a later spectral type of M2--3. The X-ray spectrum is well described by an absorbed power law with a high energy cutoff as well as by Comptonisation of soft photons from the magnetic poles of a neutron star, with heavy obscuration occurring in the surrounding medium. 
We do not detect any periodic variability with \textit{NuSTAR}, but suggest that a signal beyond 200\,s -- which is where we would expect a signal from both neutron stars accreting from late-type giants/supergiants and heavily obscured systems -- is likely present and goes undetected due to insufficient counting statistics provided by the data at low frequencies. The detection of the pulse period would require deeper X-ray observations. In general, we find the \textit{NuSTAR} lightcurves fairly stable, as they show no significant variability. Understanding of the physics here, requires comparison to lightcurves of other HMXBs at similarly stable low luminosity states, and is beyond the scope of the current study. Unveiling the geometry of the absorber would require more sensitive X-ray observations in the soft energy band that are most affected by absorption and the strong iron line, with an instrument such as \textit{XMM-Newton}, or the Resolve instrument onboard the recently launched \textit{XRISM} observatory \citep{2020arXiv200304962X}, while longer optical observations are required to study the orbital parameters of the system.

\begin{acknowledgements}\label{thanx}
  We thank the members of the XMAG collaboration for fruitful
  discussions regarding this work. We thank Amy Lien for providing us with the 157-Month \textit{Swift}/BAT lightcurve. We thank the anonymous referee for their useful comments that vastly improved this paper. 
  
  We acknowledge funding from the
  Deutsche Forschungsgemeinschaft within the eROSTEP research unit
  under DFG project number 414059771. RB and JBC acknowledge support by NASA under award number 80GSFC21M0002. CMD acknowledges support from the European Space Agency as an ESA Research Fellow. JW and ESL acknowledge support from Deutsche Forschungsgemeinschaft grant WI 1860/11-2.\\
  This work is based on data from eROSITA, the soft X-ray instrument aboard SRG, a joint Russian-German science mission supported by the Russian Space Agency (Roskosmos), in the interests of the Russian Academy of Sciences represented by its Space Research Institute (IKI), and the Deutsches Zentrum für Luft- und Raumfahrt (DLR). The SRG spacecraft was built by Lavochkin Association (NPOL) and its subcontractors, and is operated by NPOL with support from the Max Planck Institute for Extraterrestrial Physics (MPE). The development and construction of the eROSITA X-ray instrument was led by MPE, with contributions from the Dr. Karl Remeis Observatory Bamberg and ECAP (FAU Erlangen-Nuernberg), the University of Hamburg Observatory, the Leibniz Institute for Astrophysics Potsdam (AIP), and the Institute for Astronomy and Astrophysics of the University of Tübingen, with the support of DLR and the Max Planck Society. The Argelander Institute for Astronomy of the University of Bonn and the Ludwig Maximilians Universität Munich also participated in the science preparation for eROSITA. The eROSITA data shown here were processed using the eSASS/NRTA software system developed by the German eROSITA consortium. We thank the
  \textit{NuSTAR} Science Operations Centre (SOC) for their invaluable
  help in the quick scheduling of the observations. This research has made use of ISIS functions (ISISscripts) provided by ECAP/Remeis observatory and MIT (\url{https://www.sternwarte.uni-erlangen.de/isis/}). This research has
  made use of data, software and/or web tools obtained from the High
  Energy Astrophysics Science Archive Research Center (HEASARC), a
  service of the Astrophysics Science Division at NASA/GSFC and of the
  Smithsonian Astrophysical Observatory's High Energy Astrophysics
  Division. This research has made use of the Vizier and HEASARC
  database systems for querying objects and getting information from
  different catalogues. Optical analysis was carried out based on observations collected at the European Organisation for Astronomical Research in the Southern Hemisphere under ESO programmes 105.20DA.001 and 106.21B1.001, which were taken from the ESO archive. This work has made use of data from the European Space Agency (ESA) mission
\textit{Gaia} (\url{https://www.cosmos.esa.int/gaia}), processed by the \textit{Gaia} Data Processing and Analysis Consortium (DPAC, \url{https://www.cosmos.esa.int/web/gaia/dpac/consortium}). Funding for the DPAC has been provided by national institutions, in particular the institutions participating in the \textit{Gaia} Multilateral Agreement.  
SIMBAD database, operated at CDS, Strasbourg, France, was also used to get additional information. This publication also makes use of data products from the Two Micron All Sky Survey, which is a joint project of the University of Massachusetts and the Infrared Processing and Analysis Center/California Institute of Technology, funded by the National Aeronautics and Space Administration and the National Science Foundation.
\end{acknowledgements}

\bibliographystyle{aa} 
\bibliography{bibliography/bibliography}

\end{document}